\documentclass[universe,review,accept,moreauthors]{Definitions/mdpi}
\usepackage{amsmath}
\usepackage{amsfonts}
\usepackage{amssymb,bm}
\usepackage{siunitx}
\usepackage{color}

\firstpage{1}
\pubvolume{xx}
\issuenum{1}
\articlenumber{5}
\pubyear{2019}
\copyrightyear{2019}
\history{Received: date; Accepted: date; Published: date}
\Title{Casimir and Casimir-Polder Forces in Graphene Systems: Quantum Field Theoretical
Description and Thermodynamics}

\Author{Galina L. Klimchitskaya $^{1,2}$ and Vladimir M. Mostepanenko $^{1,2,3}$
}

\address{%
$^{1}$ \quad Central Astronomical Observatory at Pulkovo of the Russian Academy of Sciences, Saint Petersburg, 196140, Russia\\
$^{2}$ \quad Institute of Physics, Nanotechnology and
Telecommunications, Peter the Great Saint Petersburg
Polytechnic University, Saint Petersburg, 195251, Russia\\
$^{3}$ \quad Kazan Federal University, Kazan, 420008, Russia}

\corres{Correspondence: vmostepa@gmail.com}

\abstract{We review recent results on the low-temperature behaviors of the
Casimir-Polder and Casimir free energy an entropy for a polarizable atom
interacting with a graphene sheet and for two graphene sheets, respectively.
These results are discussed in the wide context of problems arising in the
Lifshitz theory of van der Waals and Casimir forces when it is applied to
metallic and dielectric bodies. After a brief treatment of different
approaches to theoretical description of the electromagnetic response of
graphene, we concentrate on the derivation of response function in the
framework of thermal quantum field theory in the Matsubara formulation
using the polarization tensor in (2+1)-dimensional space-time. The
asymptotic expressions for the Casimir-Polder and Casimir free energy and
entropy at low temperature, obtained with the polarization tensor, are
presented for a pristine graphene as well as for graphene sheets possessing
some nonzero energy gap $\Delta$ and chemical potential $\mu$ under different
relationships between the values of $\Delta$ and $\mu$. Along with reviewing
the results obtained in the literature, we present some new findings
concerning the case $\mu\neq 0$, $\Delta=0$. The conclusion is made that the
Lifshitz theory of the Casimir and Casimir-Polder forces in graphene systems
using the quantum field theoretical description of a pristine graphene, as
well as real graphene sheets with $\Delta>2\mu$ or $\Delta<2\mu$, is
consistent with the requirements of thermodynamics. The case of graphene
with $\Delta=2\mu\neq 0$ leads to an entropic anomaly, but is argued to be
physically unrealistic. The way to a resolution of thermodynamic problems
in the Lifshitz theory based on the results obtained for graphene is discussed.
}


\begin{document}
%%%%%%%%%%%%%%%%%%%%%%%%%%%%%%%%%%%%%%%%%%
\newcommand{\kb}{{k_{\bot}}}
\newcommand{\kkb}{{k_{\bot}^2}}
\newcommand{\tg}{{\tilde{\gamma}}}
\newcommand{\rM}{{r_{\rm TM}}}
\newcommand{\rE}{{r_{\rm TE}}}
\newcommand{\orM}{{r_{\rm TM}^{(0)}}}
\newcommand{\orE}{{r_{\rm TE}^{(0)}}}
\newcommand{\xk}{{{\rm i}\xi_l,k_{\bot}}}
\newcommand{\zy}{{({\rm i}\xi_l,k_{\bot},T)}}
\newcommand{\ozy}{{({\rm i}\xi_l,k_{\bot},0)}}
\newcommand{\oyt}{{(0,k_{\bot},T)}}
\newcommand{\ri}{{{\rm i}}}
\newcommand{\vF}{{\tilde{v}_F}}
\newcommand{\ho}{{\hbar\omega_c}}
\newcommand{\dT}{{\delta_T}}
\newcommand{\daT}{{\delta_T^{\rm impl}}}
\newcommand{\deT}{{\delta_{T\!}^{\rm expl}\,}}
\newcommand{\dbT}{{\delta_{T\!,\,l=0}^{\rm expl}\,}}
\newcommand{\dcT}{{\delta_{T\!,\,l\geqslant 1}^{\rm expl}\,}}
\newcommand{\yt}{{(k_{\bot},T,\Delta,\mu)}}
\newcommand{\yo}{{(k_{\bot},0,\Delta,\mu)}}
\newcommand{\eEt}{e^{\frac{t\Delta-2\mu}{2k_BT}}}
\newcommand{\deE}{e^{-\frac{\Delta-2\mu}{2k_BT}}}
\newcommand{\meE}{e^{-\frac{2\mu-\Delta}{2k_BT}}}

\newcommand{\ur}{{Equation~}}
\newcommand{\nur}{{Equations~}}
%%%%%%%%%%%%%%%%%%%%%%%%%%%%%%%%%%%%%%%%%%%%%%%%%%%%%%%%%%%%%%%%%%

\section{Introduction}
The attractive Casimir-Polder and Casimir forces act between an atom and an uncharged
ideal metal plane and between two parallel ideal metal planes, respectively, in vacuum
at zero temperature. These forces are entirely caused by the zero-point oscillations
of quantized electromagnetic field and depend on the Planck constant $\hbar$, speed
of light $c$, atom-plane or plane-plane separation $a$, and (in the case of the
Casimir-Polder force) on the static atomic polarizability $\alpha_o\equiv\alpha(0)$
\cite{1,2}
\begin{equation}
F_{CP}(a)=-\frac{3\alpha_0}{8\pi}\,\frac{\hbar c}{a^4}, \qquad
F_{C}(a)=-\frac{\pi^2}{240}\,\frac{\hbar c}{a^4}S,
\label{eq1a}
\end{equation}
\noindent
where $S$ is the plane area satisfying a condition $a\ll\sqrt{S}$.

The Casimir-Polder and Casimir forces generalize the familiar van der Waals force for
larger separations between interacting bodies where the retardation of electromagnetic
field becomes influential. Because of this, the forces in (\ref{eq1a}) depend on $c$,
whereas the van der Waals force depends on $\hbar$, rather than on $c$.

A unified theory of the van der Waals and Casimir forces between two plates made of
metallic or dielectric materials at any temperature was developed by Lifshitz
\cite{3,4,5}. In the framework of the Lifshitz theory, the electromagnetic field is
considered using the thermal quantum field theory in the Matsubara formulation
whereas the material properties are described classically by means of the
frequency-dependent dielectric permittivity. The van der Waals  and Casimir forces
are obtained from the Lifshitz theory in the limiting cases of short and large
separations, respectively. The Casimir-Polder force follows from the general
expression derived for two plates by rarefying    the material of one of them.
At nonzero temperature both the zero-point and thermal fluctuations of the
electromagnetic field contribute to the Casimir force. It is  important
to remember, however, that the dielectric permittivities of typical dielectric
and metallic materials are usually found \cite{6} using the Kubo formula or
kinetic theory under many assumptions with uncertain areas of application.
This may be considered as a source of potential problems when comparing the
Lifshitz theory with other fundamental theories and with the measurement data.
In recent years, the Lifshitz theory was generalized for the bodies of
arbitrary geometrical shape using the method of functional determinants
developed earlier in quantum field theory \cite{7,8,9,10}.
The developed formalism allows computation of the Casimir force between two
arbitrary bodies if the reflection amplitudes on their surfaces are available.
The latter in its turn depends on the material properties.

Although the Lifshitz theory was successfully used for many years in a more or less
qualitative manner, a comparison with precise measurements of the Casimir
interaction performed during the last 20 years revealed a problem.
The measurement data of many experiments on measuring the Casimir interaction
between metallic test bodies at room temperature excluded the theoretical
predictions if the dissipation of free electrons was taken into account in
computations (see \cite{11,12,13,14,15,16,17,18,19,20,21,22} and reviews in
\cite{23,24,25}). For dielectric test bodies, the theoretical predictions
of the Lifshitz theory were excluded by the data if computations were made
with account of the conductivity of dielectric materials at a constant current,
i.e., the dc conductivity \cite{26,27,28,29}.
If, alternatively, computations have been performed with omitted dissipation of
free electrons for metals and dc conductivity for dielectrics, the theoretical
predictions were found in good agreement with the measurement data
\cite{11,12,13,14,15,16,17,18,19,20,21,22,23,24,25,26,27,28,29,30}.
If to take into account that the dissipation of free electrons in metals and
the dc conductivity in dielectrics are the actually observed and well studied
physical effects, the experimental situation described above can be considered
as puzzling \cite{31}.

An agreement between the Lifshitz theory and some other fundamental physical
knowledge is also problematic. It turned out \cite{32,33,34,35,36,37,38}
that if the relaxation properties of free electrons in metals with perfect
crystal lattices are taken into account, the Casimir entropy calculated using
the Lifshitz theory does not vanish at zero temperature and depends on the
parameters of a system, i.e., the third law of thermodynamics (the Nernst heat
theorem) is violated \cite{39,40}. This unwanted conclusion can be avoided
if to take into account the residual relaxation at zero temperature \cite{41,42,43},
but for a perfect crystal lattice, which is the basic model of condensed matter
physics, the problem remains unresolved.  In a similar way, for dielectric
materials the Casimir entropy violates the Nernst heat theorem if the dc
conductivity is taken into account in calculations using the Lifshitz theory
\cite{44,45,46,47,47a,48}. For both metals and dielectrics, the Nernst heat
theorem is satisfied if the relaxation properties of free electrons
for metals \cite{32,33,34,35,36,37,38} and the dc
conductivity for dielectrics \cite{44,45,46,47,47a,48} are omitted in calculations
which is again puzzling.

All the preceding places strong emphasis on graphene which is a two-dimensional
monolayer of carbon atoms possessing the hexagonal crystal structure \cite{49}.
It has been shown \cite{49,50,51}
that at energies below 1--2~eV graphene possesses the linear
dispersion relation and is well described as a set of massless or very light
electronic quasiparticles. The respective field satisfies the Dirac equation
where the speed of light $c$ is replaced with the Fermi velocity
$v_F\approx c/300$. This was used to investigate many quantum field theoretical
effects arising in graphene interacting with external electromagnetic field like
the Klein paradox, particle creation from vacuum, the relativistic quantum Hall
effect etc. \cite{52,53,54,55,56,57,58,59}. There is also an extensive literature
on modeling the response functions of graphene to the electromagnetic field
using the Kubo formula, the transport theory, the two-dimensional Drude model,
the hydrodynamic model etc., and application of the obtained results for
calculation of the Casimir and Casimir-Polder forces
\cite{60,61,62,63,64,65,66,67,68,69,70,71,72,73,74,75,76,77,78,79,80,81,82,83,84}.

A distinguishing feature of graphene is, however, that its exact response function
can be found on the basis of first principles of thermal quantum field theory by
calculating the polarization tensor in (2+1)-dimensional space-time. This can be
done by using the methods of planar quantum electrodynamics developed earlier
\cite{85,86,87,88}. Although several important results on the polarization tensor
have been obtained over many years \cite{89,90,91,92}, the exact expressions
adapted for graphene at zero temperature, as well as applications to calculation
of the Casimir force were elaborated rather recently \cite{93}. The exact
expressions for the polarization tensor of graphene with nonzero energy gap
$\Delta$ (i.e., accounting for a nonzero mass of quasiparticles) and chemical
potential (i.e., admitting the presence of some doping concentration) at
nonzero temperature valid at all discrete Matsubara frequencies were derived
before long \cite{94} and applied in computations of the Casimir and Casimir-Polder
forces in different graphene systems \cite{94,95,96,97,98,99,100,101,102,103,104}.
Another representation for the polarization tensor of graphene with nonzero
$\Delta$ valid over the entire plane of complex frequencies, including the real
frequency axis, was found somewhat later \cite{105}. This representation was
generalized for the case of graphene with nonzero $\mu$ \cite{106}.
The results of \cite{105,106} have also been used in investigation of the Casimir
effect \cite{107,108,109,110,111,111a}, electrical conductivity \cite{112,113,114,115}
and reflectivity \cite{116,117,118,119,120,121} in graphene systems.
The polarization tensor for a strained graphene was also derived \cite{122}.

Taking into account that the electromagnetic response of graphene is found exactly
from the first principles of thermal quantum field theory, the question arises
on whether the theoretical predictions for the Casimir interaction in graphene
systems are in agreement with the measurement data and with the requirements of
thermodynamics. The first part of this question was answered positively.
The measurement data of the experiment \cite{123} were found in good agreement with
theoretical predictions of the Lifshitz theory \cite{124} where the reflection
coefficients were expressed via the polarization tensor of graphene.
To answer the second part of this question, one should investigate the analytic
behavior of the Casimir-Polder and Casimir free energy and entropy at arbitrarily
low temperature for both the pristine graphene and real graphene sheets
characterized by some nonzero values of the energy gap and chemical potential.
This is a complicated problem when it is considered that the polarization tensor
of graphene is a nontrivial complex-valued expression of the frequency, wave vector,
temperature, and, for the gapped and doped graphene, of the parameters $\Delta$
and $\mu$.

Below we review the results for the analytic low-temperature behavior of both the
Casimir-Polder and Casimir free energy obtained in the literature. We also derive
several new results for some specific relationships between  $\Delta$
and $\mu$ which have not been investigated so far. By and large we show that the
Lifshitz theory of the Casimir-Polder and Casimir interaction in graphene systems
using the polarization tensor is consistent with the requirements of
thermodynamics. It is also demonstrated that there is an entropic anomaly in the
case of graphene whose energy gap and chemical potential satisfy the exact
equality $\Delta=2\mu$. According to our argumentation, this case, however, does
not meet any physical situation. The results obtained for graphene are discussed
in the context of the unresolved problems arising in the Lifshitz theory for
metallic and dielectric materials and point the possible way to their resolution.

The structure of the review is the following. In Section~2, we briefly present
main expressions obtained in the Lifshitz theory and its generalizations
for the Casimir and Casimir-Polder free energy. Section~3 summarizes the
experimental and thermodynamic problems arising in the Lifshitz theory when
applied to metallic and dielectric materials. In Section~4, we list different
approaches to theoretical description of the electromagnetic response of graphene.
The quantum field theoretical description of this response by means of the
polarization tensor is presented in Section~5. The low-temperature behaviors of
the Caimir-Polder free energy and entropy for an atom interacting with a
pristine graphene and with graphene sheet possessing nonzero $\Delta$ and $\mu$
are contained in Sections 6 and 8, respectively. Sections 7 and 9 present
similar results for the Casimir free energy and entropy in the configuration
of two parallel graphene sheets. In Section~10 the reader will find a discussion
on how graphene may help to resolve thermodynamic problems of the Lifshitz
theory. Section~11 contains our conclusions.

%%%%%%%%%%%%%%%%%%%%%%%%%%%%%%%%%%%%%%%%%%%%%%%%%%%%%%%%%%%%%%%%%%%%%%
\section{The Lifshitz Theory of the Casimir and Casimir-Polder Forces and
Its Generalizations}

A unified theory of the van der Waals and Casimir forces was created by Lifshits \cite{3}
and then elaborated in more detail \cite{4,5}. In the original formulation \cite{3,4,5},
the free energy of two parallel material semispaces at temperature $T$
separated by the vacuum gap of
width $a$ in thermal equilibrium with the environment was
expressed as some functional of the frequency-dependent dielectric permittivities of
semispace materials. In more recent times, it was understood \cite{125} that the
Casimir free energy given by the Lifshitz formula for both nonmagnetic and magnetic
materials is nothing more nor less than the functional of reflection coefficients of
both propagating (on the mass shell) and evanescent (off the mass shell) electromagnetic
waves on the boundary surfaces. An explicit expression for the Casimir free energy per
unit area of the semispace outer boundary is given by \cite{3,4,5,23,24,25}
\begin{equation}
{\cal F}_C(a,T)=\frac{k_BT}{2\pi}\sum_{l=0}^{\infty}{\vphantom{\sum}}^{\!\prime}
\int_{0}^{\infty}\!\!\kb d\kb\sum_{\lambda}
\ln\left[1-r_{\lambda}^2(\ri\xi_l,\kb)e^{-2aq_l}\right].
\label{eq1}
\end{equation}
\noindent
Here,  $k_B$ is the Boltzmann constant, $\kb$ is the magnitude of the wave vector
projection on the planar structure, $\xi_l=2\pi k_BTl/\hbar$ with $l=0,\,1,\,2,\,\ldots$
are the Matsubara frequencies, $q_l^2=\kkb+\xi_l^2/c^2$, the sum in $\lambda$ is  over
the two independent polarizations of the electromagnetic field, transverse magnetic
($\lambda={\rm TM}$) and transverse electric  ($\lambda={\rm TE}$), and the prime on the
sum in $l$ divides the term with $l=0$ by two.

The reflection coefficients $r_{\lambda}$ coincide with the familiar amplitude Fresnel
reflection coefficients but are calculated at the pure imaginary Matsubara frequencies.
They are expressed via the dielectric permittivity $\varepsilon(\omega)$ and magnetic
permeability $\mu(\omega)$ of the semispace material in the following way:
\begin{equation}
\rM(\ri\xi_l,\kb)=\frac{\varepsilon(\ri\xi_l)q_l-k_l}{\varepsilon(\ri\xi_l)q_l+k_l},
\qquad
\rE(\ri\xi_l,\kb)=\frac{\mu(\ri\xi_l)q_l-k_l}{\mu(\ri\xi_l)q_l+k_l},
\label{eq2}
\end{equation}
\noindent
where
\begin{equation}
k_l^2=\kkb+\varepsilon(\ri\xi_l)\mu(\ri\xi_l)\frac{\xi_l^2}{c^2}.
\label{eq3}
\end{equation}

It was also understood that the Lifshitz formula for two plates of finite thickness   or
for an arbitrary number of planar material layers in place of each plate has also the
form (\ref{eq1}) if the reflection coefficients (\ref{eq2}) are replaced with with their
generalizations. For instance, for the plates consisting of any number of material layers
the coefficients $r_{\lambda}$ are expressed via the Fresnel coefficients (\ref{eq2})
by a more complicated recursion formulas \cite{126,127}. In all these cases, however,
the reflection coefficients are eventually expressed via the frequency-dependent dielectric
permittivities of plate materials.

As mentioned in Section~1, by now the Lifshitz formula for the Casimir free energy is
generalized for the case of two compact bodies of arbitrary shape \cite{7,8,9,10}.
Specifically, for two parallel planar structures of sufficiently large area $S$
separated by a distance $a\ll\sqrt{S}$ kept in thermal equilibrium at temperature $T$, it
is again given by (\ref{eq1}) where the reflection coefficients depend on the physical
nature of a planar structure. These coefficients may have little in common with the
Fresnel reflection coefficients (\ref{eq2}) if the planar structure is not at all
characterized by the concept of a dielectric permittivity or is described by a spatially
nonlocal permittivity.

The Lifshitz formula for the Casimir-Polder free energy between an atom above a planar
structure can be obtained by considering the Casimir free energy of this planar structure
interacting with a semispace described by the frequency-dependent dielectric
permittivity and Fresnel reflection coefficients (\ref{eq2}). For this purpose, the
semispace matter should be rarified and its dielectric permittivity and reflection
coefficients expanded in powers of a small number of atoms per unit volume.
Such a procedure was first performed for the case of two semispaces \cite{4,5} and
resulted in the following formula for the Casimir-Polder free energy:
\begin{equation}
{\cal F}_{CP}(a,T)=-k_BT\sum_{l=0}^{\infty}{\vphantom{\sum}}^{\!\prime}
\alpha(\ri\xi_l)
\int_{0}^{\infty}\!\!\kb d\kb q_le^{-2aq_l}
\left[\left(2-\frac{\xi_l^2}{q_l^2c^2}\right)\rM(\xk)-
\frac{\xi_l^2}{q_l^2c^2}\rE(\xk)
\right],
\label{eq4}
\end{equation}
\noindent
where $\alpha(\omega)$ is the dynamic polarizability of an atom in its ground state and
the reflection coefficients are now not necessarily
the Fresnel ones but associated with the planar
structure. The alternative approaches to a derivation of the Casimir-Polder free energy
are based on the models of a harmonic oscillator \cite{128} and point dipoles
interacting through the electromagnetic field \cite{129}.

%%%%%%%%%%%%%%%%%%%%%%%%%%%%%%%%%%%%%%%%%%%%%%%%%%%%%%%%%%%%%%%%%%%%%%%%%
\section{Experimental and Thermodynamic Problems of the Lifshitz Theory}

Experiments on measuring the Casimir force are usually performed in the
sphere-plate configuration. As an illustration of disagreement between
the measurement data and theoretical predictions of the Lifshitz theory,
mentioned in Section~1, we present two examples from the experiments
using metallic test bodies.

The Casimir force acting between a sphere of radius $R$ and a plate, both
coated with metallic films, can be calculated in the proximity force
approximation \cite{24} as
\begin{equation}
{F}_C^{\,sp}(a,T)=2\pi R{\cal F}_C(a,T),
\label{eq5}
\end{equation}
\noindent
where the Casimir free energy for two parallel plates is defined in (\ref{eq1}).
Note that at $a\ll R$ the exact expression for ${F}_C^{\,sp}$, obtained using
generalizations of the Lifshitz theory, deviate from (\ref{eq5}) by only a
fraction of a percent \cite{130,131,132,133,134}.

In Figure~1, we present typical results from the most recent experiment \cite{22},
where the gradient ${{F}_C^{\,sp}}^{\,\prime}$ of the force (\ref{eq5}) was
measured between the Au-coated test bodies.
%%%%%%%%%%%%%%%%%%%%%%%%%%%%%%%%%
%%%__Figure_1__%%%%%%%%%%%%%%%%%%
%%%%%%%%%%%%%%%%%%%%%%%%%%%%%%%%
\begin{figure}[!ht]
\centering
\vspace*{-2.5cm}
\includegraphics[width=16 cm]{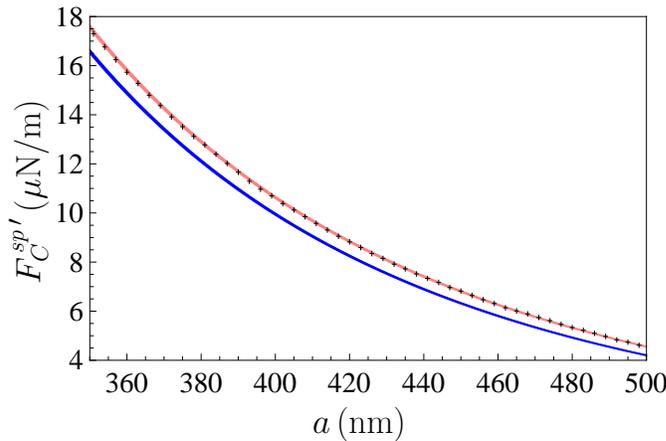}
\vspace*{-14cm}
\caption{The mean measured gradients of the Casimir force between
an Au-coated sphere and an Au-coated plate
are shown by crosses as a function of separations. For clarity only
each third experimental data point is plotted.
The bottom and top lines demonstrate theoretical predictions
of the Lifshitz theory obtained  with inclusion and neglect of the
relaxation of conduction electrons, respectively.\label{fig1}}
\end{figure}
%%%%%%%%%%%%%%%%%%%%%%%%%%%%%%%%
In this figure, the experimental data with their errors are shown as crosses.
The bottom line shows the predictions of the Lifshitz theory obtained by
using the seemingly most reliable dielectric permittivity of Au. Its behavior
at relatively high frequency is found from the available optical data for
the complex index of refraction of Au which is extrapolated to lower
frequencies by the well tested Drude model taking into account the dissipation
of free electrons. The behavior of $\varepsilon$ along the imaginary frequency
axis is obtained using the Kramers-Kronig relation (see in \cite{23,24} for
more details). The top line in Figure~1 shows the predictions of the Lifshitz
theory obtained if the dielectric permittivity of Au given by the optical data
is extrapolated to lower frequencies by the plasma model which completely
disregards the relaxation properties of free electrons.
From Figure~1 it is clearly seen that the top line is in a good agreement
with the measurement data, whereas the bottom line is excluded by them.
This result is puzzling because the plasma model is in fact applicable only at
high frequencies in the region of infrared optics, where the dissipation of free
electrons does not play any role. So, in our case it was used outside of its
validity region. All sets of data in the experiment \cite{22} excluded the
Lifshitz theory taking the dissipation of free electrons into account
over the separation range from 250~nm to $1.1~\mu$m.

One more representative example on the disagreement of a literally understood
Lifshitz theory with the measurement data is shown in Figure~2.
%%%%%%%%%%%%%%%%%%%%%%%%%%%%%%%%%
%%%__Figure_1__%%%%%%%%%%%%%%%%%%
%%%%%%%%%%%%%%%%%%%%%%%%%%%%%%%%
\begin{figure}[!b]
\centering
\vspace*{-2.5cm}
\includegraphics[width=16 cm]{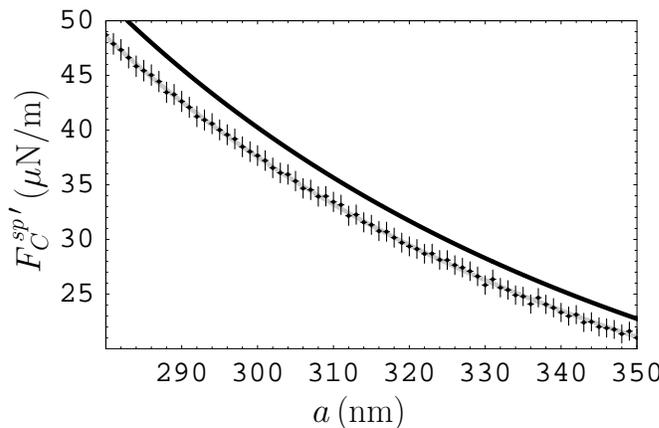}
\vspace*{-14cm}
\caption{The mean measured gradients of the Casimir force between  a Ni-coated
sphere and a Ni-coated plate
are shown by crosses as a function of separations.
The top and bottom lines demonstrate theoretical predictions
of the Lifshitz theory obtained with inclusion and neglect of the
relaxation of conduction electrons, respectively.\label{fig1}}
\end{figure}
%%%%%%%%%%%%%%%%%%%%%%%%%%%%%%%%
Here, the gradient of the Casimir force between a sphere and a plate coated with the
films of a magnetic metal Ni was measured \cite{17,18}. The theoretical predictions
obtained with included dissipation of free electrons (i.e., using the Drude model at
low frequencies) are shown by the top line, whereas the predictions found with this
dissipation omitted (i.e., using the plasma model at
low frequencies) are shown by the bottom line. We emphasize that for magnetic metals
these lines exchange their places as compared to the nonmagnetic ones in Figure~1.
As is seen in Figure~2, the theoretical predictions with taken into account dissipation
of free electrons are excluded by the data. At the same time, the predictions of the
Lifshitz theory disregarding this dissipation are in a good agreement with the data.
We recall that similar results have been obtained in a number of experiments for
metallic \cite{11,12,13,14,15,16,17,18,19,20,21,22} and dielectric \cite{26,27,28,29,30}
test bodies with up to a factor of 1000 differences between two alternative
theoretical predictions in the experiment \cite{19} using the differential force
measurement scheme \cite{135,136,137}.

Now we briefly discuss the thermodynamic problems faced by the Lifshitz theory. From
Equation~(\ref{eq1}) one can easily find the Casimir entropy per unit area of the plates
defined as
\begin{equation}
S_C(a,T)=-\frac{\partial{\cal F}_C(a,T)}{\partial T}.
\label{eq6}
\end{equation}

For two semispaces made of a nonmagnetic metal with perfect crystal lattices it was proven
that the Casimir entropy at zero temperature calculated using the Drude model with account
of dissipation of free electrons takes the following value \cite{32,33,34}:
\begin{equation}
S_C(a,0)=-\frac{k_B\zeta(3)}{16\pi a^2}\left[1-
4\frac{c}{a\omega_p}+12\left(\frac{c}{a\omega_p}\right)^2-\ldots\right]< 0,
\label{eq7}
\end{equation}
\noindent
where $\zeta(z)$ is the Riemann zeta function and $\omega_p$ is the plasma frequency.

It is seen that this value depends on the parameters of a system, such as $a$ and
$\omega_p$, and, thus, the third law of thermodynamics (the Nernst heat theorem) is
violated in this case \cite{39,40}. If, however, the dielectric permittivity of the
plasma model is used, which disregards the dissipation of free electrons, one obtains
\cite{32,33,34}
\begin{equation}
S_C(a,0)=0,
\label{eq8}
\end{equation}
\noindent
i.e., the Nernst heat theorem is satisfied.  Similar results hold for other Casimir
configurations containing metallic test bodies made of both nonmagnetic and magnetic
metals \cite{35,36,37}.

For dielectric semispaces the Casimir entropy at zero temperature calculated with
taken into account conductivity at a constant current has the form \cite{44,45,46,47}
\begin{equation}
S_C(a,T)=\frac{k_B}{16\pi a^2}\left\{\zeta(3)-{\rm Li}_3\left[\left(
\frac{\varepsilon(0)-1}{\varepsilon(0)+1}\right)^2\right]\right\}>0,
\label{eq9}
\end{equation}
\noindent
where ${\rm Li}_n(z)$ is the polylogarithm function and $\varepsilon(0)$ is the static value
of the dielectric permittivity. The right-hand side of this equation again depends on the
parameters of a system which means a violation of the Nernst heat theorem.
If, however, the dc conductivity is omitted in calculation, one obtains the zero value
(\ref{eq8}) of the Casimir entropy at zero temperature,
i.e., the Nernst heat theorem is satisfied.

For the Casimir-Polder entropy, defined as
\begin{equation}
S_{CP}(a,T)=-\frac{\partial{\cal F}_{CP}(a,T)}{\partial T},
\label{eq10}
\end{equation}
\noindent
where ${\cal F}_{CP}$ is given in Equation~(\ref{eq4}), the situation is somewhat
different. The point is that both contradictions of the Lifshitz theory with  the
measurement data and thermodynamics for metallic test bodies are caused by the TE
contribution to the Matsubara terms with $l=0$ in \ur(\ref{eq1}).
In the  Casimir-Polder free energy (\ref{eq4}), however, the reflection coefficient
$\rE(0,\kb)$ enters with a factor $\xi_0^2=0$ and, thus, does not contribute
to the result. That is why the Lifshitz theory of the Casimir-Polder interaction
of a polarizable atom with a metallic plate is in agreement with thermodynamics
even if the low-frequency electromagnetic response of a metal is described by
the Drude model.

A different situation arises for an atom interacting  with a dielectric plate.
In this case the Casimir-Polder entropy at zero temperature calculated with
taken into account conductivity of a plate material at a constant current
(dc conductivity) is given by \cite{45,47a}
\begin{equation}
S_{CP}(a,0)=\frac{k_B}{4a^3}\alpha(0)
\frac{\varepsilon(0)-1}{\varepsilon(0)+1},
\label{eq11}
\end{equation}
\noindent
where $\alpha(0)$ is the static atomic polarizability. Thus, the Nernst heat theorem
is violated. Later this result was generalized for a magnetizable atom and
a ferromagnetic dielectric plate \cite{48}.
If, however, the dc conductivity of a plate material is omitted in calculations,
one obtains in place of (\ref{eq11})
\begin{equation}
S_{CP}(a,0)=0,
\label{eq12}
\end{equation}
\noindent
i.e., the Nernst heat theorem is satisfied \cite{45,47a,48}.
We emphasize that in the single precise experiment on measuring the Casimir-Polder
force at separations of a few micrometers \cite{30}, where the difference between
alternative theoretical predictions of the Lifshitz theory is relatively large, the
measurement data were found in agreement with theory disregarding the dc conductivity
\cite{30} and exclude the theory taking the dc conductivity into account \cite{27}.
This situation again presents a puzzle or, as it is also called, a conundrum \cite{111a}
because the dc conductivity of dielectrics at nonzero temperature is an observable
and well studied physical effect, so that there is no reason why one should omit it
when comparing experiment with theory.

In the end of this section, we note that the violation of the Nernst heat theorem
and the disagreement between the measured Casimir and Casimir-Polder forces and
predictions of the Lifshitz theory with included dc conductivity
of a dielectric plate are again caused
by the term with $l=0$ in \nur(\ref{eq1}) and (\ref{eq4}). Here, however, this
happens due to the TM contribution to  \nur(\ref{eq1}) and (\ref{eq4})
as opposed to the case of Casimir interaction between two metallic plates where
similar effects are caused by the TE contribution to  \ur(\ref{eq1}).

%%%%%%%%%%%%%%%%%%%%%%%%%%%%%%%%%%%%%%%%%%%%%%%%%%%%%%%%%%%%%%%%%%%%%%%%%%
\section{Different Approaches to Theoretical Description of the
Electromagnetic Response of Graphene}

As was explained in Section~2, the Casimir interaction between two planar structures,
as well as between an atom and a planar structure, can be described by \nur(\ref{eq1})
and (\ref{eq4}), respectively, if the reflection coefficients on a planar structure
are defined appropriately. Thus, \nur(\ref{eq1}) and (\ref{eq4}) are applicable
for theoretical description of the Casimir interaction between two graphene sheets,
as well as of the Casimir-Polder interaction between an atom and a graphene sheet.
This raises a question of what is the form of the reflection coefficients on a
graphene sheet. The point is that the standard Fresnel coefficients (\ref{eq2}) or
their generalizations for the plates of finite thickness \cite{24} are expressed
via the frequency-dependent dielectric permittivity which is well defined only for
volumetric bodies containing sufficiently large number of atomic layers.
This condition is not satisfied for graphene which is a one-atom thick layer of
carbon atoms \cite{49,50,51}.

As a first approximation, graphene was treated as a two-dimensional free-electron gas
characterized by some typical wave number $K=6.75\times 10^5~\mbox{m}^{-1}$ which is
determined by the hexagonal structure of graphite \cite{67,68}. The reflection
coefficients of the electromagnetic waves on such a sheet take the form \cite{67,68,138}
\begin{equation}
\rM(\xk)=\frac{c^2q_lK}{c^2q_lK+\xi_l^2}, \qquad
\rE(\xk)=-\frac{K}{K+q_l}.
\label{eq13}
\end{equation}
\noindent
\nur(\ref{eq1}) and (\ref{eq4}) with the reflection coefficients (\ref{eq13}) were used
to calculate the Casimir and Casimir-Polder interactions in various carbon nanostructures
\cite{65,67,68,69,70,138,139}. It should be taken into account, however, that the
plasma-sheet approach, which is also often called "the hydrodynamic model", fails to
account for a linearity of the dispersion relation for graphene. Finally, it was
demonstrated that the theoretical predictions of this approach are in contradiction
\cite{140} with the experimental data on measuring the Casimir force in graphene systems
\cite{123}.

In the widely used approach, the reflection coefficients on a graphene sheet are expressed
via the longitudinal, $\chi^{\|}(\xk)$, and transverse, $\chi^{\bot}(\xk)$,
density-density correlation functions of graphene \cite{82,141,142}
\begin{equation}
\rM(\xk)=\frac{2\pi e^2q_l\chi^{\|}(\xk)}{2\pi e^2q_l\chi^{\|}(\xk)-\kkb},
\qquad
\rE(\xk)=-\frac{2\pi e^2\xi_l^2\chi^{\bot}(\xk)}{2\pi e^2\xi_l^2\chi^{\bot}(\xk)-
c^2\kkb q_l}.
\label{eq14}
\end{equation}
\noindent
The density-density correlation functions are connected with the nonlocal electric
susceptibilities (polarizabilities) of graphene by
$\chi^{\|,\bot}=-\kb\alpha^{\|,\bot}/(2\pi e^2)$.
The density-density correlation functions have been used in computations of the
Casimir force \cite{63,76,79,80,82}.
It should be noted, however, that although the formulas (\ref{eq14}) are exact,
the exact expressions for the correlation functions $\chi^{\|,\bot}$ at nonzero temperature
remain unknown with exception of only some particular cases.

In a similar way, the reflection coefficients on a graphene sheet were expressed via
the in-plane, $\sigma^{\|}(\xk)$, and out-of-plane, $\sigma^{\bot}(\xk)$, electrical
conductivities of graphene \cite{79,82}
\begin{equation}
\rM(\xk)=\frac{2\pi q_l\sigma^{\|}(\xk)}{2\pi q_l\sigma^{\|}(\xk)+\xi_l},
\qquad
\rE(\xk)=-\frac{2\pi \xi_l\sigma^{\bot}(\xk)}{2\pi \xi_l\sigma^{\bot}(\xk)+
c^2q_l}.
\label{eq15}
\end{equation}
\noindent
These expressipons are also exact because the conductivities are expressed via the
density-density correlation functions as
$\sigma^{\|,\bot}=-e^2\xi_l\chi^{\|,\bot}/\kb^2$ \cite{82}.
In doing so, the exact expressions for the conductivities of graphene at nonzero
temperature also remained unknown. Thus, the Casimir force between two graphene sheets
was computed under an assumption that the graphene conductivity can be modeled by the
in-plane optical conductivity of graphite with no account \cite{77} or with account
\cite{81} of spatial dispersion. This modeling includes the two-dimensional Drude
term and a series of Lorentz oscillators.

By and large several approximate methods for calculation of the Casimir interaction
have been elaborated but their accuracy and the region of applicability remained
uncertain.

%%%%%%%%%%%%%%%%%%%%%%%%%%%%%%%%%%%%%%%%%%%%%%%%%%%%%%%%%%%%%%%%%%%%%%%%%%%%
\section{Quantum Field Theoretical Description of the Electromagnetic Response of
Graphene via the Polarization Tensor}

As mentioned in Section~1, the field of massless or having some small mass $m$
electronic excitations in graphene satisfy the three-dimensional Dirac equation
where the speed of light $c$ is replaced with the Fermi velocity $v_F$.
For graphene possessing the chemical potential $\mu$ this equation takes the
form \cite{94,143}
\begin{equation}
\left[\frac{\gamma^0}{v_F}\left(\ri\hbar\frac{\partial}{\partial t}-\mu\right)+
\ri\hbar\left(\gamma^1\frac{\partial}{\partial x^1}+
\gamma^2\frac{\partial}{\partial x^2}\right)-mv_F\right]\psi(x)=0,
\label{eq16}
\end{equation}
\noindent
where the Dirac matrices satisfy the conditions
$(\gamma^0)^2=-(\gamma^{1,2})^2=1$.

Multiplying this equation by the dimensionless Fermi velocity $\vF=v_F/c\approx 1/300$,
we can rewrite  it in the form
\begin{equation}
\left[\frac{\tg^0}{c}\left(\ri\hbar\frac{\partial}{\partial t}-\mu\right)+
\ri\hbar\left(\tg^1\frac{\partial}{\partial x^1}+
\tg^2\frac{\partial}{\partial x^2}\right)-mc\vF^2\right]\psi(x)=0,
\label{eq17}
\end{equation}
\noindent
where $\tg^0=\gamma^0$ and  $\tg^{1,2}=\vF\gamma^{1,2}$.
Note that $2mv_F^2=\Delta$, where $\Delta$ is the energy gap, so that
$mc\vF^2=\Delta/(2c)$.
Then, an interaction with the electromagnetic field $A_{\beta}(x)$ is introduced by the
replacement
\begin{equation}
\ri\hbar\frac{\partial}{\partial x^{\beta}}\to
\ri\hbar\frac{\partial}{\partial x^{\beta}}-\frac{e}{c}A_{\beta}(x),
\label{eq18}
\end{equation}
\noindent
where $x^{\beta}=(ct,x^1,x^2)$. As a result, the Dirac equation (\ref{eq17}) is
given by
\begin{equation}
\left[\tg^0\left(\ri\hbar\frac{\partial}{\partial t}-eA_0-\mu\right)+
\tg^1\left(\ri\hbar c\frac{\partial}{\partial x^1}-eA_1\right)+
\tg^2\left(\ri\hbar c\frac{\partial}{\partial x^2}-eA_2\right)
-\frac{\Delta}{2}\right]\psi(x)=0.
\label{eq19}
\end{equation}

In the one-loop approximation, an interaction of the graphene quasiparticles with
the electromagnetic field is described by the diagram consisting of an electronic
quasiparticle loop with two photon legs. At zero temperature this corresponds
to the following expression for the polarization tensor \cite{94,144}
\begin{equation}
\Pi^{\beta\delta}(k_0,\mbox{\boldmath$k_{\bot}$})=\ri\frac{e^2}{c}\int
\frac{dq_0\,d^2\mbox{\boldmath$q_{\bot}$}}{(2\pi)^3}
{\rm tr}\left[S(q_0,\mbox{\boldmath$q_{\bot}$})\tg^{\beta}
S(q_0-k_0,\mbox{\boldmath$q_{\bot}$}-\mbox{\boldmath$k_{\bot}$})\tg^{\delta}\right],
\label{eq20}
\end{equation}
\noindent
where $(q_0,\mbox{\boldmath$q_{\bot}$})$ and $(k_0,\mbox{\boldmath$k_{\bot}$})$
are the three-dimensional wave vectors of a loop electronic excitation and
an external photon, respectively, and the propagator of the quasiparticles takes
the form \cite{94}
\begin{equation}
S(q_0,\mbox{\boldmath$q_{\bot}$})=-\frac{\tg^0[q_0+\mu/(\hbar c)]-
\tg^1q_1-\tg^2q_2-\Delta/(2\hbar c)}{[q_0+\mu/(\hbar c)+
\ri\,\epsilon\,{\rm sign}q_0]^2-\vF^2{\mbox{\boldmath$q$}_{\bot}^2}-
\Delta^2/(4\hbar^2c^2)}.
\label{eq21}
\end{equation}

For taking into account nonzero temperature in the framework of the Matsubara formalism,
one should replace an integration with respect to $q_0$ in (\ref{eq20}) with
a summation over the pure imaginary fermionic Matsubara frequencies given by
\begin{equation}
cq_{0n}=2\pi\ri\left(n+\frac{1}{2}\right)\frac{k_BT}{\hbar},
\label{eq22}
\end{equation}
\noindent
where $n=\pm 1,\,\pm 2,\,\ldots\,$. One should also put $k_{0l}=\ri\xi_l/c$,
where $\xi_l$ are the bosonic Matsubara frequencies. Taking into account that
there are four fermionic species for graphene, the polarization tensor (\ref{eq20})
is expressed as
\begin{equation}
\Pi^{\beta\delta}(k_{0l},\mbox{\boldmath$k_{\bot}$})=-8\pi\alpha k_BT
\sum_{n=-\infty}^{\infty}\left(n+\frac{1}{2}\right)
\int\frac{\,d^2\mbox{\boldmath$q_{\bot}$}}{(2\pi)^2}
{\rm tr}\left[S(q_{0n},\mbox{\boldmath$q_{\bot}$})\tg^{\beta}
S(q_{0n}-k_{0l},\mbox{\boldmath$q_{\bot}$}-\mbox{\boldmath$k_{\bot}$})\tg^{\delta}\right].
\label{eq23}
\end{equation}

Explicit expressions for this tensor which can be analytically continued over the
entire plane of complex frequencies were found in \cite{105,106}.
They are the functions of the Matsubara frequencies, magnitude of the photon wave
vector, temperature, energy gap and chemical potential. In fact the polarization tensor
 in a medium is completely defined by the two independent quantities, e.g., $\Pi_{00}$
 and $\Pi_{\beta}^{\,\beta}$ \cite{94}. For our purposes, instead of a trace, it is
 more convenient to exploit
 \begin{equation}
 \Pi(\xk)\equiv\kkb\Pi_{\beta}^{\,\beta}(\xk)-q_l^2\Pi_{00}(\xk)
\label{eq24}
\end{equation}
\noindent
as the second quantity. Below we use the abbreviate notations
\begin{equation}
\Pi_{00}(\xk)\equiv\Pi_{00,l}, \qquad
\Pi(\xk)\equiv\Pi_{l},
\label{eq25}
\end{equation}
\noindent
and write out all or some of the arguments $\xk,T,\Delta$, and $\mu$ where it is
helpful for better understanding.

We present the explicit expressions for the following two contributions to
$\Pi_{00,l}$ and $\Pi_l$:
\begin{eqnarray}
&&
\Pi_{00}(\xk,T,\Delta,\mu)=\Pi_{00}^{(0)}(\xk,\Delta)+
\Pi_{00}^{(1)}(\xk,T,\Delta,\mu),
\nonumber \\
&&
\Pi(\xk,T,\Delta,\mu)=\Pi^{(0)}(\xk,\Delta)+
\Pi^{(1)}(\xk,T,\Delta,\mu).
\label{eq26}
\end{eqnarray}
\noindent
Here, the quantities $\Pi_{00}^{(0)}$ and $\Pi^{(0)}$ describe the undoped graphene
with $\mu=0$ but, generally speaking, $\Delta\neq 0$ at zero temperature.
They are given by \cite{93}
\begin{equation}
\Pi_{00,l}^{(0)}=\frac{\alpha\hbar c{\kkb}}{p_l}\Psi(D_l), \qquad
\Pi_{l}^{(0)}=\alpha\hbar{\kkb}\frac{p_l}{c}\Psi(D_l),
\label{eq27}
\end{equation}
\noindent
where $p_l^2=v_F^2{\kkb}+\xi_l^2$, $D_l=\Delta/(\hbar p_l)$ and
\begin{equation}
\Psi(x)=2\left[x+(1-x^2)\arctan(x^{-1})\right].
\label{eq28}
\end{equation}

The quantities $\Pi_{00}^{(1)}$ and $\Pi^{(1)}$ in (\ref{eq26}) describe both
the thermal effect and the dependence on $\mu$. They can be presented in the
form \cite{105,106,110}
\begin{eqnarray}
&&
\Pi_{00,l}^{(1)}=\frac{4\alpha\hbar c p_l}{v_F^2}\int_{D_l}^{\infty}\!\!\!du
\left(\frac{1}{e^{B_lu+\frac{\mu}{k_BT}}+1}+\frac{1}{e^{B_lu-\frac{\mu}{k_BT}}+1}\right)
\nonumber \\
&&~~~~~~~~~~~~~~~~~~~~~~~~~~~~~~
\times\left[1-{\rm Re}\frac{1-u^2+2\ri\frac{\xi_l}{p_l}u}{\left(1-u^2+2\ri\frac{\xi_l}{p_l}u
+\frac{v_F^2{\kkb}}{p_l^2}D_l^2\right)^{1/2}}\right],
\nonumber \\
&&
\Pi_{l}^{(1)}=-\frac{4\alpha\hbar p_l\xi_l^2}{cv_F^2}\int_{D_l}^{\infty}\!\!\!du
\left(\frac{1}{e^{B_lu+\frac{\mu}{k_BT}}+1}+\frac{1}{e^{B_lu-\frac{\mu}{k_BT}}+1}\right)
\nonumber \\
&&~~~~~~~~~~~~~~~~~~~~~~~~~~~~~~
\times\left[1-{\rm Re}\frac{1-\frac{p_l^2}{\xi_l^2}u^2+2\ri\frac{p_l}{\xi_l}u+
\frac{v_F^2{\kkb}}{\xi_l^2}D_l^2}{\left(1-u^2+2\ri\frac{\xi_l}{p_l}u
+\frac{v_F^2{\kkb}}{p_l^2}D_l^2\right)^{1/2}}\right],
\label{eq29}
\end{eqnarray}
\noindent
where $B_l=\hbar p_l/(2k_BT)$.

In order to investigate one specific case, which was not considered in the literature
so far, we also need the expressions for $\Pi_{00,l}$ and $\Pi_l$ at $T=0$ calculated
under the condition $\Delta<2\mu$. They were obtained in \cite{110} by using
\nur(\ref{eq26})--(\ref{eq28}) and (\ref{eq29}) taken at $T=0$. The result is
\begin{eqnarray}
&&
\Pi_{00}(\xk,0,\Delta,\mu)=\frac{8\alpha c\mu}{v_F^2}-
\frac{\alpha c\hbar{\kkb}}{p_l}\left\{2M_l{\rm Im}(y_l\sqrt{1+y_l^2})
\right.
\nonumber \\
&&~~~~~~~~~~~~~~~~~~~~~~~~~~~~~~
\left.
+
(2-M_l)\left[2{\rm Im}\ln(y_l+\sqrt{1+y_l^2})-\pi\right]\right\},
\nonumber \\
&&
\Pi(\xk,0,\Delta,\mu)=-\frac{8\alpha\xi_l^2\mu}{cv_F^2}+
\frac{\alpha \hbar p_l{\kkb}}{c}\left\{2M_l{\rm Im}(y_l\sqrt{1+y_l^2})
\right.
\nonumber \\
&&~~~~~~~~~~~~~~~~~~~~~~~~~~~~~~
\left.
-
(2-M_l)\left[2{\rm Im}\ln(y_l+\sqrt{1+y_l^2})-\pi\right]\right\},
\label{eq30}
\end{eqnarray}
\noindent
where the following notations are introduced
\begin{equation}
M_l=1+D_l^2,\qquad
y_l\equiv y_l(\Delta,\mu)=\frac{\hbar\xi_l+2\ri\mu}{\hbar v_F\kb\sqrt{M_l}}.
\label{eq31}
\end{equation}

The reflection coefficients on a graphene sheet can be expressed in terms of
the polarization tensor given by
\nur(\ref{eq26})--(\ref{eq29}) \cite{94}
\begin{equation}
\rM(\xk)=\frac{q_l\Pi_{00}(\xk)}{q_l\Pi_{00}(\xk)+2\hbar{\kkb}},
\qquad
\rE(\xk)=-\frac{\Pi(\xk)}{\Pi(\xk)+2\hbar{\kkb}q_l}.
\label{eq32}
\end{equation}

Taking into account a connection between the polarization tensor and
density-density correlation functions \cite{104}
\begin{equation}
\chi^{\|}(\xk)=-\frac{1}{4\pi e^2\hbar}\Pi_{00}(\xk),
\qquad
\chi^{\bot}(\xk)=-\frac{c^2}{4\pi e^2\hbar\xi_l^2}\Pi(\xk),
\label{eq33a}
\end{equation}
\noindent
the reflection coefficients (\ref{eq32}) become equivalent to (\ref{eq14}).
Thus, calculation of explicit expressions (\ref{eq26})--(\ref{eq29}) for
the polarization tensor of graphene at any $T$ and with any values of
$\Delta$ and $\mu$ on the basis of first principles of thermal quantum field
theory resulted in respective expressions for the density-density correlation
functions which were previously available only in some specific cases.

\nur(\ref{eq26})--(\ref{eq29}) have been used in calculations of the Casimir
interaction in graphene systems \cite{107,108,109,110,111,111a}.
After an analytic continuation to the real frequency axis, they were also
applied for investigation of the electrical conductivity \cite{112,113,114,115}
and reflectances of graphene \cite{116,117,118,119,120,121}.
 This allows to compare the longitudinal and transverse dielectric
permittivities of graphene $\varepsilon^{\|,\bot}$, which are directly expressed
via the respective conductivities $\sigma^{\|,\bot}$, with the dielectric
permittivity of the Drude model $\varepsilon_D$. As shown in \cite{114}, for
$\hbar\omega<\Delta$ it holds  ${\rm Re}\,\sigma^{\|,\bot}=0$.
This results in ${\rm Im}\,\varepsilon^{\|,\bot}=0$ unlike the Drude model for
which ${\rm Im}\,\varepsilon_D\sim 1/\omega$ when $\omega$ goes to zero, but
similar to the plasma model for which ${\rm Im}\,\varepsilon_p=0$.
The imaginary part of $\sigma^{\|,\bot}$ for graphene is contained in equation (47)
of \cite{114}. In the asymptotic limit $\omega\to 0$, for both cases
$\Delta>2\mu$ and $\Delta<2\mu$ it was shown that
${\rm Im}\,\sigma^{\|,\bot}\sim 1/\omega$ (see equations (50) and (54) in
\cite{114}) which results in
 ${\rm Re}\,\varepsilon^{\|,\bot}\sim -1/\omega^2$.
 This is again unlike the Drude model for which ${\rm Re}\,\varepsilon_D$
 takes a finite value at $\omega=0$, but in direct analogy to the plasma
 model which does not take into account the relaxation properties of free
 electrons (see Section 3). Thus, the dielectric properties of graphene,
 expressed via the polarization tensor, provide reason enough to guess
 that the Casimir entropy calculated using the reflection coefficients
(\ref{eq32}) satisfies the Nernst heat theorem like it holds for the
reflection coefficients (\ref{eq2}) and the dielectric permittivity of
the plasma model.

Below we discuss an application of the polarization tensor for a treatment of
thermodynamic properties of the Casimir-Polder and Casimir interactions in
graphene systems and make sure that this guess is correct.

%%%%%%%%%%%%%%%%%%%%%%%%%%%%%%%%%%%%%%%%%%%%%%%%%%%%%%%%%%%%%%%%%%%%%%%%%%%%
\section{Low-Temperature Behavior of the Casimir-Polder Free Energy and Entropy
for an Atom and a Pristine Graphene Sheet}

The asymptotic behavior of the free energy for an atom interacting with a sheet
of ideal graphene possessing $\Delta=\mu=0$ at low $T$ was investigated using the
formalism of the polarization tensor \cite{145}.
The Casimir-Polder free energy was expressed according to (\ref{eq4}) and the
reflection coefficients presented in (\ref{eq32}). In this case, the
polarization tensor was significantly simplified by putting $\Delta=\mu=0$
in \nur(\ref{eq26})--(\ref{eq31}).

The main idea of the perturbation approach used in \cite{145} is to present the
polarization tensor as a sum of two contributions (\ref{eq26}), where for
a pristine graphene with  $\Delta=\mu=0$ the quantities $\Pi_{00,l}^{(1)}$ and
$\Pi_{l}^{(1)}$ have the meaning of the thermal corrections, and to make an
expansion in powers of the small parameters
\begin{equation}
\frac{\Pi_{00,l}^{(1)}}{\Pi_{00,l}^{(0)}}\ll 1, \qquad
\frac{\Pi_{l}^{(1)}}{\Pi_{l}^{(0)}}\ll 1.
\label{eq33}
\end{equation}
\noindent
The smallness of these parameters is guaranteed by the fact that their numerators
go to zero with vanishing temperature, whereas denominators are equal to the
nonvanishing quantities $\Pi_{00}\ozy$ and $\Pi\ozy$ in the case of a pristine
graphene.

As a result, the reflection coefficients (\ref{eq32}) are presented as a sum of
the contributions at zero temperature and the thermal corrections to them
\begin{equation}
r_{\rm TM,TE}(\xk)=r_{\rm TM,TE}^{(0)}(\xk)+\dT r_{\rm TM,TE}(\xk).
\label{eq34}
\end{equation}
\noindent
Here, the first contributions on the right-hand side are given by (\ref{eq32})
where, instead $\Pi_{00,l}$ and $\Pi_l$, we substitute $\Pi_{00,l}^{(0)}$ and
$\Pi_l^{(0)}$ defined in ({\ref{eq27}) but with $\Delta=0$:
\begin{equation}
\orM(\xk)=\frac{\alpha\pi cq_l}{\alpha\pi cq_l+2p_l}, \qquad
\orE(\xk)=-\frac{\alpha\pi p_l}{\alpha\pi p_l+2cq_l}.
\label{eq35}
\end{equation}
\noindent
The second contributions to (\ref{eq34}) are expressed as
\begin{equation}
\dT\rM(\xk)=\frac{2\alpha\pi cq_lp_l}{(\alpha\pi cq_l+2p_l)^2}\,
\frac{\Pi_{00,l}^{(1)}}{\Pi_{00,l}^{(0)}},
\qquad
\dT\rE(\xk)=-\frac{2\alpha\pi cq_lp_l}{(\alpha\pi p_l+2cq_l)^2}\,
\frac{\Pi_{l}^{(1)}}{\Pi_{l}^{(0)}},
\label{eq36}
\end{equation}
\noindent
where $\Pi_{00,l}^{(1)}$ and $\Pi_l^{(1)}$ are defined by ({\ref{eq29})
with $\Delta=\mu=0$.

Now we note that the temperature-dependent part of the Casimir-Polder
free energy can be presented as
\begin{equation}
\dT{\cal F}_{CP}(a,T)={\cal F}_{CP}(a,T)-E_{CP}(a),
\label{eq37}
\end{equation}
\noindent
where ${\cal F}_{CP}$ is presented in (\ref{eq4}) and the Casimir-Polder energy
at $T=0$ is given by
\begin{equation}
{E}_{CP}(a)=-\frac{\hbar}{2\pi}\int_{0}^{\infty}\!\!\!d\xi
\alpha(\ri\xi)
\int_{0}^{\infty}\!\!\kb d\kb qe^{-2aq}
\left[\left(2-\frac{\xi^2}{q^2c^2}\right)\orM(\ri\xi,\kb)-
\frac{\xi^2}{q^2c^2}\orE(\ri\xi,\kb)
\right],
\label{eq38}
\end{equation}
\noindent
where $q^2={\kkb}+\xi^2/c^2$.

Substituting \ur(\ref{eq34}) in the Casimir-Polder free energy (\ref{eq4}),
the thermal correction to the Casimir-Polder energy $E_{CP}$ can be
identically presented as
\begin{equation}
\dT{\cal F}_{CP}(a,T)=\daT{\cal F}_{CP}(a,T)+\deT{\cal F}_{CP}(a,T),
\label{eq39}
\end{equation}
\noindent
where the contributions to the right-hand side of this equation are defined as
\begin{eqnarray}
&&
\daT{\cal F}_{CP}(a,T)=-k_BT\sum_{l=0}^{\infty}{\vphantom{\sum}}^{\!\prime}
\alpha(\ri\xi_l)
\int_{0}^{\infty}\!\!\kb d\kb q_le^{-2aq_l}
\nonumber \\
&&~~~~
\times
\left[\left(2-\frac{\xi_l^2}{q_l^2c^2}\right)\orM(\xk)-
\frac{\xi_l^2}{q_l^2c^2}\orE(\xk)
\right]
-E_{CP}(a)
\label{eq40}
\end{eqnarray}
\noindent
and
\begin{eqnarray}
&&
\deT{\cal F}_{CP}(a,T)=-k_BT\sum_{l=0}^{\infty}{\vphantom{\sum}}^{\!\prime}
\alpha(\ri\xi_l)
\int_{0}^{\infty}\!\!\kb d\kb q_le^{-2aq_l}
\nonumber \\
&&~~~~~~~~~~~~~~~
\times
\left[\left(2-\frac{\xi_l^2}{q_l^2c^2}\right)\dT\rM(\xk)-
\frac{\xi_l^2}{q_l^2c^2}\dT\rE(\xk)
\right].
\label{eq41}
\end{eqnarray}
\noindent

Here, $\daT{\cal F}_{CP}$ contains the reflection coefficients at $T=0$, so that its
dependence on $T$ originates from only the Matsubara summation. This is the
reason why it is called "implicit".  The contribution  $\deT{\cal F}_{CP}$ would be
equal to zero for the reflection coefficients independent on $T$ as a parameter.
Because of this it called "explicit".

It is convenient also to split $\deT{\cal F}_{CP}$ in two parts
\begin{equation}
\deT{\cal F}_{CP}(a,T)=\dbT{\cal F}_{CP}(a,T)+\dcT{\cal F}_{CP}(a,T),
\label{eq42}
\end{equation}
\noindent
where the first part is equal to the term of (\ref{eq41}) with $l=0$ and the
second part is equal to the sum of all remaining terms with $l\geqslant 1$.

As is seen from \nur(\ref{eq38}) and (\ref{eq40}), the implicit thermal
correction $\daT{\cal F}_{CP}$ is represented by a difference between the sum
and the integral which can be calculated by using the Abel-Plana formula
\cite{24,146}
\begin{equation}
\sum_{l=0}^{\infty}{\vphantom{\sum}}^{\!\prime}F(l)-
\int_{0}^{\infty}\!\!\!F(t)dt=\ri\int_{0}^{\infty}\!
\frac{F(\ri t)-F(-\ri t)}{e^{2\pi t}-1}dt,
\label{eq43}
\end{equation}
\noindent
which is valid for a function $F(z)$ analytic in the right half-plane.

Using this approach, it was shown that at sufficiently low temperature
satisfying the condition $k_BT\ll\hbar v_F/(2a)$ the implicit thermal
correction decreases with temperature as (see \cite{145} for details)
\begin{equation}
\daT{\cal F}_{CP}(a,T)\sim -\frac{\alpha_0(k_BT)^4}{(\hbar c)^3},
\label{eq44}
\end{equation}
\noindent
where $\alpha_0=\alpha(0)$ is the static atomic polarizability.
Here and below we preserve only dimensional parameters in the asymptotic
formulas.  This behavior is determined by the TM contribution to the
Casimir-Polder free energy. The result (\ref{eq44}) is similar to that for an atom
interacting with a dielectric plate \cite{24,45}.

Now we deal with the first part of the explicit thermal correction,
$\dbT{\cal F}_{CP}$, given by the term with $l=0$ in (\ref{eq41}).
It is determined by only the TM mode because $\xi_0=0$.
In this case the derivation of the asymptotic expression at low $T$
results in \cite{145}
\begin{equation}
\dbT{\cal F}_{CP}(a,T)\sim \frac{\alpha_0(k_BT)^4}{(\hbar c)^3}
\ln\frac{ak_BT}{\hbar v_F}.
\label{eq45}
\end{equation}
\noindent
One can see that (\ref{eq45}) becomes greater in magnitude that the
magnitude of (\ref{eq44}) with decreasing $T$.

The last contribution to the thermal correction is the second part of
(\ref{eq42}), $\dcT{\cal F}_{CP}$, which is equal to the sum of all terms
of (\ref{eq41}) with $l\geqslant 1$. The asymptotic behavior of this part
at low $T$ can be found taking into account that the major contribution
to the integral with respect to $\kb$ in (\ref{eq41}) is given by $\kb$
satisfying the condition $q_l\sim 1/(2a)$. Then at sufficiently low
temperature $k_BT\ll\hbar v_F/(2a)$, one arrives at \cite{145}
\begin{equation}
\dcT{\cal F}_{CP}(a,T)\sim -\frac{\alpha_0(k_BT)^3}{v_F^2\hbar^2a}.
\label{eq46}
\end{equation}
\noindent
This goes to zero slower than (\ref{eq44}) and (\ref{eq45}) and is
again determined by the TM reflection coefficient.

One can conclude that for an atom interacting with a pristine graphene
sheet the thermal correction (\ref{eq37}) to the Casimir-Polder energy
behaves at low temperature as
\begin{equation}
\dT{\cal F}_{CP}(a,T)\sim
\dcT{\cal F}_{CP}(a,T)\sim -\frac{\alpha_0(k_BT)^3}{v_F^2\hbar^2a}.
\label{eq47}
\end{equation}
\noindent

The resulting Casimir-Polder entropy vanishes with vanishing temperature
by the power law
\begin{equation}
S_{CP}(a,T)\sim \frac{\alpha_0k_B(k_BT)^2}{v_F^2\hbar^2a}.
\label{eq48}
\end{equation}
\noindent
This means that the Lifshitz theory of atom-graphene interaction using the
reflection coefficients expressed via the polarization tensor satisfies the
requirements of thermodynamics for a pristine graphene.

The asymptotic results (\ref{eq47}) and (\ref{eq48}) obtained for a pristine
graphene are presented in column 2 of Table~1 (lines 2 and 3, respectively).
%%%%%%%%%%__Table__1__%%%%%%%
%%%%%%%%%%%%%%%%%%%%%%%%%%%%%%%%%
\begin{table}[!ht]
 \centering
\tablesize{\normalsize}
 \caption{Up to an order of magnitude asymptotic behaviors at arbitrarily low
temperature of the thermal corrections to the Casimir-Polder energy (line 2)
and of the Casimir-Polder entropy (line 3) under different relationships
between the energy gap $\Delta$ and chemical potential $\mu$ (line 1).}
 \begin{tabular}{ccccc}
  \toprule
 & $\Delta=\mu=0$ & $\Delta>2\mu\geqslant 0$& $\Delta=2\mu\neq0$ &  $0\leqslant\Delta<2\mu$\\
 \midrule
$\dT{\cal F}_{CP}(a,T)$  & $-\frac{\alpha_0(k_BT)^3}{v_F^2\hbar^2a}$ &
$-\frac{\alpha_0(k_BT)^5}{(\hbar c)^3\Delta}$ & $-\frac{\alpha_0k_BT}{a^3}$ &
$-\frac{\alpha_0\mu^2(k_BT)^2}{(\hbar c)^2a\sqrt{4\mu^2-\Delta^2}}$\\
\midrule
$S_{CP}(a,T)$ & $\frac{\alpha_0k_B(k_BT)^2}{v_F^2\hbar^2a}$ &
$\frac{\alpha_0k_B(k_BT)^4}{(\hbar c)^3\Delta}$ & $\frac{\alpha_0k_B}{a^3}$ &
$\frac{\alpha_0\mu^2k_B^2T}{(\hbar c)^2a\sqrt{4\mu^2-\Delta^2}}$\\
  \bottomrule
 \end{tabular}
\end{table}
%%%%%%%%%%%%%%%%%%%%%%%%%%%%%%%%%%

%%%%%%%%%%%%%%%%%%%%%%%%%%%%%%%%%%%%%%%%%%%%%%%%%%%%%%%%%%%%%%%%%%%%%%%%%%%%
\section{Low-Temperature Behavior of the Casimir Free Energy and Entropy
for Two Pristine Graphene Sheets}

Two parallel graphene sheets separated by some distance $a$ most closely resemble the
original Casimir configuration of two ideal metal planes \cite{1}.
An investigation of the thermodynamic properties of the Casimir interaction between
two graphene sheets is a more complicated problem than for an atom interacting with
graphene because the Lifshitz formula (\ref{eq1}) is a nonlinear function of the
reflection coefficients in contrast to the Lifshitz formula (\ref{eq4}) describing
the Casimir-Polder interaction. Nevertheless, this problem has been solved \cite{147}
following the same approach as presented in Section~6 for an atom interacting with
a pristine graphene sheet.  Below we briefly summarize the obtained results.

Similar to \ur(\ref{eq37}), we define the thermal correction to the Casimir energy
$E_C(a)$ per unit area of two graphene sheets at zero temperature
\begin{equation}
\dT{\cal F}_{C}(a,T)={\cal F}_{C}(a,T)-E_{C}(a),
\label{eq49}
\end{equation}
\noindent
where the Casimir free energy ${\cal F}_{C}$ is defined in (\ref{eq1}) and
$E_C$ is given by
\begin{equation}
{E}_C(a)=\frac{\hbar}{4\pi^2}\int_{0}^{\infty}\!\!\!\!d\xi
\int_{0}^{\infty}\!\!\kb d\kb\sum_{\lambda}
\ln\left[1-{r_{\lambda}^{(0)}}^2(\ri\xi,\kb)e^{-2aq}\right].
\label{eq50}
\end{equation}
\noindent
Here, the reflection coefficients at $T=0$, $r_{\lambda}^{(0)}$, are
presented in (\ref{eq35}).

By adding and subtracting from (\ref{eq49}) the quantity having the same form as
the Casimir free energy (\ref{eq1}), but containing the zero-temperature
reflection coefficients $r_{\lambda}^{(0)}$ in place of $r_{\lambda}$, one rearrange
(\ref{eq49}) to the following equivalent form:
\begin{equation}
\dT{\cal F}_{C}(a,T)=\daT{\cal F}_{C}(a,T)+\deT{\cal F}_{C}(a,T),
\label{eq51}
\end{equation}
\noindent
where
\begin{equation}
\daT{\cal F}_C(a,T)=\frac{k_BT}{2\pi}\sum_{l=0}^{\infty}{\vphantom{\sum}}^{\!\prime}
\int_{0}^{\infty}\!\!\kb d\kb\sum_{\lambda}
\ln\left[1-{r_{\lambda}^{(0)}}^2(\xk)e^{-2aq_l}\right]-E_C(a).
\label{eq52}
\end{equation}
\noindent
and
\begin{equation}
\deT{\cal F}_C(a,T)={\cal F}_C(a,T)-
\frac{k_BT}{2\pi}\sum_{l=0}^{\infty}{\vphantom{\sum}}^{\!\prime}
\int_{0}^{\infty}\!\!\kb d\kb\sum_{\lambda}
\ln\left[1-{r_{\lambda}^{(0)}}^2(\xk)e^{-2aq_l}\right].
\label{eq53}
\end{equation}
\noindent
In doing so, one obtains for two graphene sheets the same separation of the
thermal correction in two contributions as was obtained in (\ref{eq39}) for
the Casimir-Polder interaction. According to (\ref{eq52}), the implicit
contribution $\daT{\cal F}$ depends on temperature only through the
Matsubara summation, whereas the contribution $\deT{\cal F}$, defined in
(\ref{eq53}), vanishes if there is no explicit dependence of the reflection
coefficients on temperature as a parameter.

The implicit contribution (\ref{eq52}) to the thermal correction is defined
as a difference between the sum and the integral and can be calculated
using the Abel-Plana formula (\ref{eq43}) similar to the case of the
Casimir-Polder interaction considered in Section~6. Calculations were
performed under the condition $k_BT\ll\hbar v_F/(2a)$ and taking into
account that the major contribution to the integral (\ref{eq43}) is given
by $t\sim 1/(2\pi)$. According to the obtained results (see \cite{147}
for details),
\begin{equation}
\daT{\cal F}_C(a,T)\sim -\frac{(k_BT)^3}{(\hbar v_F)^2}.
\label{eq54}
\end{equation}
\noindent
This low-temperature behavior is determined by the TM contribution to
\ur(\ref{eq52}). Note that the TE contribution is of the same form as
(\ref{eq54}) and has an opposite sign, but is much smaller than the
magnitude of (\ref{eq54}).

In order to find the low-temperature behavior of the explicit thermal correction,
we substitute (\ref{eq34}) in (\ref{eq53}) and use the following evident identity:
\begin{eqnarray}
&&
\ln\left\{1-\left[r_{\lambda}^{(0)}(\xk)+\dT r_{\lambda}(\xk)\right]^2
e^{-2aq_l}\right\}-\ln\left[1-{r_{\lambda}^{(0)}}^2(\xk)e^{-2aq_l}\right]
\nonumber \\
&&
=\ln\left\{1-\frac{2r_{\lambda}^{(0)}(\xk)\dT r_{\lambda}(\xk)+
[\dT r_{\lambda}(\xk)]^2}{1-{r_{\lambda}^{(0)}}^2(\xk)e^{-2aq_l}}
e^{-2aq_l}\right\}.
\label{eq55}
\end{eqnarray}
\noindent
Taking into account that $\dT r_{\lambda}$ goes to zero with vanishing $T$,
one can expand the logarithm in (\ref{eq55}) up to the first order in this
small quantity. Then, at sufficiently low temperature, \ur(\ref{eq53}) can
be rewritten as
\begin{equation}
\deT{\cal F}_C(a,T)=-\frac{k_BT}{\pi}\sum_{l=0}^{\infty}{\vphantom{\sum}}^{\!\prime}
\int_{0}^{\infty}\!\!\kb d\kb e^{-2aq_l}\sum_{\lambda}
\frac{r_{\lambda}^{(0)}(\xk)\dT r_{\lambda}(\xk)}{1-
{r_{\lambda}^{(0)}}^2(\xk)e^{-2aq_l}}.
\label{eq56}
\end{equation}

The reflection coefficients $r_{\lambda}^{(0)}$ at $T=0$ entering this equation
are defined in (\ref{eq35}) and the thermal corrections to them in (\ref{eq36}).
Note that, similar to (\ref{eq42}), the explicit thermal correction is
conveniently presented as
\begin{equation}
\deT{\cal F}_{C}(a,T)=\dbT{\cal F}_{C}(a,T)+\dcT{\cal F}_{C}(a,T),
\label{eq57}
\end{equation}
\noindent
where $\dbT{\cal F}_{C}$ is equal to the term of (\ref{eq56}) with $l=0$ and
$\dcT{\cal F}_{C}$ --- to the sum of all terms in (\ref{eq56}) with
$l\geqslant 1$.

A derivation of the asymptotic behavior of both $\dbT{\cal F}_{C}$ and
$\dcT{\cal F}_{C}$ at arbitrarily low temperature was performed in \cite{147}.
Here we present only the main results. Thus, under the condition
$k_BT\ll\hbar v_F/(2a)$ one finds
\begin{equation}
\dbT{\cal F}_C(a,T)\sim -\frac{(k_BT)^3}{(\hbar v_F)^2},
\label{eq58}
\end{equation}
\noindent
i.e., the same behavior as in (\ref{eq54}), which is determined by the TM
contribution to (\ref{eq56}). The TE contribution is again positive and
much smaller than the magnitude of (\ref{eq58}).

For the sum of all Matsubara terms  with $l\geqslant 1$ in (\ref{eq56}),
similar analytic derivation results in the following behavior
at arbitrarily low temperature (see \cite{147} for details):
\begin{equation}
\dcT{\cal F}_C(a,T)\sim \frac{(k_BT)^3}{(\hbar c)^2}
\ln\frac{ak_BT}{\hbar c}.
\label{eq59}
\end{equation}
\noindent
This up to an order of magnitude estimation is valid for both the TM and
TE contributions. However, the numerical coefficient in front of the TE
contribution is by the factor of $10^{-8}$ smaller than in front of the
TM one \cite{147}. Thus, the correction $\dcT{\cal F}_C$ is again
determined by the TM mode of the electromagnetic field.

By comparing \nur(\ref{eq54}), (\ref{eq58}), and (\ref{eq59}), one
concludes that the major contribution to the total thermal correction
(\ref{eq51}) at low temperature is given by $\dcT{\cal F}_C$, i.e.,
\begin{equation}
\dT{\cal F}_C(a,T)\sim
\dcT{\cal F}_C(a,T)\sim \frac{(k_BT)^3}{(\hbar c)^2}
\ln\frac{ak_BT}{\hbar c}.
\label{eq60}
\end{equation}
\noindent
Similar to the case of the Casimir-Polder interaction in (\ref{eq47}),
this correction is negative. The respective Casimir entropy per unit area
of graphene sheet
\begin{equation}
S_C(a,T)=-\frac{\partial\dT{\cal F}_C(a,T)}{\partial T}
\sim -k_B\left(\frac{k_BT}{\hbar c}\right)^2
\ln\frac{ak_BT}{\hbar c}
\label{eq61}
\end{equation}
\noindent
is positive and goes to zero with vanishing temperature. Thus, the Lifshitz
theory of the Casimir interaction between two pristine graphene sheets is
thermodynamically consistent if the electromagnetic response of graphene
is described by the polarization tensor.

In Table~2, the asymptotic results (\ref{eq60}) and (\ref{eq61}) obtained for a pristine
graphene are included in column 2  (lines 2 and 3, respectively).
%%%%%%%%%%__Table__2__%%%%%%%
%%%%%%%%%%%%%%%%%%%%%%%%%%%%%%%%%
\begin{table}[!ht]
 \centering
\tablesize{\normalsize}
 \caption{Up to an order of magnitude asymptotic behaviors at arbitrarily low
temperature of the thermal corrections to the Casimir energy (line 2) and of
the Casimir entropy (line 3) under different relationships between the
energy gap $\Delta$ and chemical potential $\mu$ (line 1).}
 \begin{tabular}{ccccc}
  \toprule
 & $\Delta=\mu=0$ & $\Delta>2\mu\geqslant 0$& $\Delta=2\mu\neq0$ &  $0\leqslant\Delta<2\mu$\\
 \midrule
$\dT{\cal F}_{C}(a,T)$  & $\frac{(k_BT)^3}{(\hbar c)^2}\ln\frac{ak_BT}{\hbar c}$ &
$-\frac{(k_BT)^5}{(\hbar c)^2\Delta^2}$ & $-\frac{k_BT}{a^2}$ &
$-\frac{a(4\mu^2-\Delta^2)(k_BT)^2}{(\hbar c)^3}$\\
\midrule
$S_{C}(a,T)$ & $-k_B\left(\frac{k_BT}{\hbar c}\right)^2\ln\frac{ak_BT}{\hbar c}$ &
$\frac{k_B(k_BT)^4}{(\hbar c)^2\Delta^2}$ & $\frac{k_B}{a^2}$ &
$\frac{a(4\mu^2-\Delta^2)k_B^2T}{(\hbar c)^3}$\\
  \bottomrule
 \end{tabular}
\end{table}
%%%%%%%%%%%%%%%%%%%%%%%%%%%%%%%%%%

%%%%%%%%%%%%%%%%%%%%%%%%%%%%%%%%%%%%%%%%%%%%%%%%%%%%%%%%%%%%%%%%%%%%%%%%%%%%
\section{Low-Temperature Behavior of the Casimir-Polder Free Energy and Entropy
for an Atom and a Graphene Sheet Possessing the Energy Gap and Chemical Potential}

Now we consider the free energy for an atom interacting with real graphene sheet
characterized by some values of $\Delta$ and $\mu$. In this case one should
use the full expressions (\ref{eq26})--(\ref{eq29}) for the polarization tensor
valid for any $\Delta$ and $\mu$. Moreover, in this case, depending on the
relationship between the values of $\Delta$ and $2\mu$, the quantities
$\Pi_{00,l}^{(1)}$ and $\Pi_{l}^{(1)}$ may not have a meaning of the thermal
corrections to the polarization tensor at zero temperature.
Below we present the results obtained in the literature \cite{148} and consider
one more case which was not investigated so far.

We start with graphene possessing a relatively small chemical potential
satisfying the condition $\Delta>2\mu$. This case is somewhat similar to that
of a pristine graphene because the quantities $\Pi_{00,l}^{(0)}$ and $\Pi_{l}^{(0)}$
defined in (\ref{eq27}) have the meaning of the polarization tensor components
at $T=0$
\begin{equation}
\Pi_{00,l}^{(0)}=\Pi_{00}(\xk,0,\Delta), \qquad
\Pi_{l}^{(0)}=\Pi(\xk,0,\Delta),
\label{eq62}
\end{equation}
\noindent
whereas the quantities $\Pi_{00,l}^{(0)}$ and $\Pi_{l}^{(0)}$
defined in (\ref{eq29}) are the thermal corrections to them
\begin{equation}
\Pi_{00,l}^{(1)}=\dT\Pi_{00}(\xk,T,\Delta,\mu), \qquad
\Pi_{l}^{(1)}=\dT\Pi(\xk,T,\Delta,\mu).
\label{eq62}
\end{equation}
\noindent
In accordance with this, under the condition $\Delta>2\mu$ the polarization tensor
at $T=0$ does not depend on $\mu$, the quantities (\ref{eq62})  defined in (\ref{eq27})
do not vanish,  and the perturbation theory in the parameters (\ref{eq33}) is
applicable.

The thermal correction to the Casimir-Polder energy at $T=0$ is again presented
by \nur(\ref{eq39})--(\ref{eq41}) as a sum of the implicit and explicit contributions.
The low-temperature behavior of an implicit contribution is found using the
Abel-Plana formula (\ref{eq43}) under an assumption $\Delta/(\hbar\omega_c)>1$,
which is applicable at sufficiently large atom-graphene separations.
The result is (see \cite{148} for details)
\begin{equation}
\daT{\cal F}_{CP}(a,T)\sim -\frac{\alpha_0(k_BT)^5}{(\hbar c)^3\Delta}.
\label{eq64}
\end{equation}
\noindent
{}From this equation it is seen that $\daT{\cal F}_{CP}$ for a real graphene sheet
vanishes with temperature faster than the respective result (\ref{eq44}) obtained
for a pristine graphene.

The explicit contribution to the thermal correction in (\ref{eq39}) is given by
(\ref{eq41}) and is again presented as the sum of two parts (\ref{eq42}).
Under the conditions $\Delta>2\mu$, $k_BT\ll\Delta-2\mu$, and
$\Delta/(\hbar\omega_c)>1$ the first part is estimated as \cite{148}
\begin{equation}
\dbT{\cal F}_{CP}(a,T)\sim -\frac{\alpha_0(k_BT)^2}{a^2\hbar c}\deE.
\label{eq65}
\end{equation}
%\noindent

Under the same conditions, the low-temperature behavior of the second part of
explicit thermal correction is given by \cite{148}
\begin{equation}
\dcT{\cal F}_{CP}(a,T)\sim -\frac{\alpha_0k_BT}{a^3}\deE.
\label{eq66}
\end{equation}
%\noindent

As a result, for a real graphene sheet with $\Delta>2\mu$ the total thermal
correction to the Casimir-Polder energy decreases with $T$ by the following law:
\begin{equation}
\dT{\cal F}_{CP}(a,T)\sim
\daT{\cal F}_{CP}(a,T)\sim -\frac{\alpha_0(k_BT)^5}{(\hbar c)^3\Delta}.
\label{eq67}
\end{equation}
%\noindent

The respective Casimir-Polder entropy behaves as
\begin{equation}
S_{CP}(a,T)\sim \frac{\alpha_0k_B(k_BT)^4}{(\hbar c)^3\Delta}
\label{eq68}
\end{equation}
\noindent
and goes to zero with vanishing temperature in accordance to the
Nernst heat theorem.

The asymptotic results (\ref{eq67}) and (\ref{eq68}) for
graphene with $\Delta>2\mu$ are obtained using the expansions in three small
parameters
\begin{equation}
\frac{4\pi k_BTa}{\hbar c}\ll 1, \qquad
\frac{\hbar v_F}{2a\Delta}\ll 1, \qquad
e^{-\frac{\Delta-2\mu}{2k_BT}}\ll 1.
\label{eq68a}
\end{equation}
\noindent
These results
are included in column 3 of Table~1 (lines 2 and 3, respectively).

As shown in \cite{148}, the results obtained for the case $\Delta>2\mu$
can be analytically continued for graphene with $\Delta=2\mu$.
Because of this, \nur(\ref{eq64})-- (\ref{eq66}) are also applicable in this
 case leading, however, to quite a different conclusion.
Although (\ref{eq64}) remains unchanged for $\Delta=2\mu$, \nur(\ref{eq65})
and (\ref{eq66}) reduce to
\begin{equation}
\dbT{\cal F}_{CP}(a,T)\sim -\frac{\alpha_0(k_BT)^2}{a^2\hbar c},
\qquad
\dcT{\cal F}_{CP}(a,T)\sim -\frac{\alpha_0k_BT}{a^3}.
\label{eq69}
\end{equation}
\noindent
Thus, for graphene with $\Delta=2\mu$ the leading behavior of the thermal
correction at low $T$ is given not by (\ref{eq64}), but by the second
equation in (\ref{eq69}):
\begin{equation}
\dT{\cal F}_{CP}(a,T)\sim
\dcT{\cal F}_{CP}(a,T)\sim -\frac{\alpha_0k_BT}{a^3}.
\label{eq70}
\end{equation}
%\noindent

The respective Casimir-Polder entropy
\begin{equation}
S_{CP}(a,T)\sim \frac{\alpha_0k_B}{a^3}
\label{eq71}
\end{equation}
\noindent
does not vanish at zero temperature and depends on the parameters of a system
which means a violation of the Nernst heat theorem. The physical meaning
of this violation is discussed in Section~10.

The results for graphene with $\Delta=2\mu$ were obtained using the asymptotic
expansions in only the first two small parameters indicated in (\ref{eq68a}).
In Table~1, these  results
are included in column 4 (lines 2 and 3, respectively).

We note that \nur(\ref{eq64})-- (\ref{eq68}) are applicable to graphene
with any chemical potential satisfying the condition $\Delta>2\mu$
and, specifically, for graphene with $\mu=0$.  In so doing the energy gap
$\Delta$ is not equal to zero. As to \nur(\ref{eq69})-- (\ref{eq71}),
they are valid for graphene with $\Delta=2\mu\neq 0$ because the condition
$\Delta\neq 0$ was assumed in the derivation of the low-temperature
behavior. Hence, the limiting transition to the case of a pristine
graphene in these equations is impossible.

The last case to consider is the Casimir-Polder interaction between an atom
and doped graphene sheet with relatively large chemical potential
satisfying the condition $\Delta<2\mu$. This case is the most complicated because
the quantities $\Pi_{00,l}^{(0)}$ and  $\Pi_{l}^{(0)}$ in (\ref{eq27}) are no
longer the components of the polarization tensor at $T=0$ [the latter depend
on $\mu$ and are defined in \ur(\ref{eq30})]. In a similar way,
the quantities $\Pi_{00,l}^{(1)}$ and  $\Pi_{l}^{(1)}$ in (\ref{eq29}) are no
longer the  thermal corrections to them.

The thermal correction $\daT{\cal F}_{CP}$ can be calculated by using its
definition (\ref{eq40}) where the reflection coefficients are defined by
(\ref{eq32}) taken at $T=0$ and expressed via the polarization tensor
(\ref{eq30}) depending on $\mu$
\begin{eqnarray}
&&
\orM(\xk)=
\frac{q_l\Pi_{00,l}(\kb,0,\Delta,\mu)}{q_l\Pi_{00,l}(\kb,0,\Delta,\mu)+2\hbar{\kkb}},
\nonumber \\
&&
\orE(\xk)=-\frac{\Pi_l(\kb,0,\Delta,\mu)}{\Pi_l(\kb,0,\Delta,\mu)+2\hbar{\kkb}q_l}.
\label{eq72}
\end{eqnarray}
\noindent
Then, the low-temperature behavior of $\daT{\cal F}_{CP}$ can be found by using
the Abel-Plana formula (\ref{eq43}) under the conditions
$\sqrt{4\mu^2-\Delta^2}>\hbar\omega_c$, which is valid at sufficiently large
atom-graphene separations, and $k_BT\ll\hbar\omega_c$. The result is
(see \cite{148} for details)
\begin{equation}
\daT{\cal F}_{CP}(a,T)\sim
-\frac{\alpha_0\mu^2(k_BT)^2}{(\hbar c)^2a\sqrt{4\mu^2-\Delta^2}}.
\label{eq73}
\end{equation}

The explicit thermal correction to the Casimir-Polder energy is given by (\ref{eq41}),
where the thermal correction to the reflection coefficient $\dT\rM$ can be obtained
in the first order of the parameter
\begin{equation}
\frac{\dT\Pi_{00,l}(\kb,T,\Delta,\mu)}{\Pi_{00,l}(\kb,0,\Delta,\mu)}\ll 1.
\label{eq74}
\end{equation}
\noindent
Here, in accordance to (\ref{eq30}), $\Pi_{00,l}(\kb,0,\Delta,\mu)\neq 0$ at any $l$.
Substituting (\ref{eq34}) in the first equation of (\ref{eq32}), one obtains
\begin{equation}
\dT\rM(\xk)=
\frac{2\hbar q_l{\kkb}\dT\Pi_{00,l}(\kb,T,\Delta,\mu)}{\left[
q_l\Pi_{00,l}(\kb,0,\Delta,\mu)+2\hbar{\kkb}\right]^2}\,,
\label{eq75}
\end{equation}
\noindent
where
\begin{equation}
\dT\Pi_{00,l}(\kb,T,\Delta,\mu)\equiv\Pi_{00,l}^{(1)}(\kb,T,\Delta,\mu)
-\Pi_{00,l}^{(1)}(\kb,0,\Delta,\mu)
\label{eq76}
\end{equation}
\noindent
and $\Pi_{00,l}^{(1)}$ is defined in (\ref{eq29}).

For finding $\dbT{\cal F}_{CP}(a,T)$, there is no need in $\dT\rE(0,\kb)$ because
in accordance to (\ref{eq41}) it does not contribute to this part of the thermal
correction. As a result, under the conditions $k_BT\ll\hbar\omega_c$ and
$\Delta>\hbar\omega_c$ the asymptotic behavior of our interest takes the form
\cite{148}
\begin{equation}
\dbT{\cal F}_{CP}(a,T)\sim \frac{\alpha_0k_BT}{a^3}\meE.
\label{eq77}
\end{equation}

The part of the thermal correction $\dcT{\cal F}_{CP}$, in accordance to
(\ref{eq41}), depends on both $\dT\rM$ and  $\dT\rE$.
By coincidence, just for $l\geqslant 1$ it holds $\Pi_{l}(\kb,0,\Delta,\mu)\neq 0$
and the perturbation theory in the small parameter
\begin{equation}
\frac{\dT\Pi_{l}(\kb,T,\Delta,\mu)}{\Pi_{l}(\kb,0,\Delta,\mu)}\ll 1
\label{eq78}
\end{equation}
\noindent
is applicable, where, similar to (\ref{eq76}),
\begin{equation}
\dT\Pi_{l}(\kb,T,\Delta,\mu)\equiv\Pi_{l}^{(1)}(\kb,T,\Delta,\mu)
-\Pi_{l}^{(1)}(\kb,0,\Delta,\mu)
\label{eq78a}
\end{equation}
\noindent
and $\Pi_{l}^{(1)}$ is defined in (\ref{eq29}).

Substituting (\ref{eq34}) in the second equality in (\ref{eq32}),
we have
\begin{equation}
\dT\rE(\xk)=
-\frac{2\hbar q_l{\kkb}\dT\Pi_{l}(\kb,T,\Delta,\mu)}{\left[
\Pi_{l}(\kb,0,\Delta,\mu)+2\hbar q_l{\kkb}\right]^2}\, .
\label{eq79}
\end{equation}
\noindent
Note that for $l=0$ one has $\Pi_{0}(\kb,0,\Delta,\mu)=0$ and the perturbation
theory in the parameter (\ref{eq78}) becomes inapplicable. This case, however,
is irrelevant to the Casimir-Polder interaction due to the factor $\xi_0^2=0$
in front of  $\dT\rE(0,\kb)$ in (\ref{eq41}) but is of importance for two
doped graphene sheets considered in Section~9.

Using the formulas outlined above, the low-temperature behavior of the last
contribution to the thermal correction can be found under the conditions
$k_BT\ll 2\mu-\Delta$, $\sqrt{4\mu^2-\Delta^2}>\hbar\omega_c$ and
$\Delta>\hbar\omega_c$ with the result \cite{148}
\begin{equation}
\dcT{\cal F}_{CP}(a,T)\sim \frac{\alpha_0\hbar c}{a^4}\meE.
\label{eq80}
\end{equation}

By comparing \nur(\ref{eq73}), (\ref{eq77}), and (\ref{eq80}), one concludes
that the low-temperature behavior of the total thermal correction to the
Casimir-Polder energy for an atom and graphene with $\Delta<2\mu$ is given by
\begin{equation}
\dT{\cal F}_{CP}(a,T)\sim\daT{\cal F}_{CP}(a,T)\sim
-\frac{\alpha_0\mu^2(k_BT)^2}{(\hbar c)^2a\sqrt{4\mu^2-\Delta^2}}.
\label{eq81}
\end{equation}

The respective Casimir-Polder entropy at low temperature behaves as
\begin{equation}
S_{CP}(a,T)\sim
\frac{\alpha_0\mu^2k_B^2T}{(\hbar c)^2a\sqrt{4\mu^2-\Delta^2}},
\label{eq82}
\end{equation}
\noindent
i.e., vanishes when $t$ goes to zero in accordance to the Nernst heat theorem.

\nur(\ref{eq81}) and (\ref{eq82}) were obtained by using the first two small
parameters presented in (\ref{eq68a}), whereas the third one was replaced with
\begin{equation}
e^{-\frac{2\mu-\Delta}{2k_BT}}\ll 1.
\label{eq82a}
\end{equation}
\noindent
The asymptotic results (\ref{eq81}) and (\ref{eq82}) valid for $\Delta<2\mu$
(including $\Delta=0$, see below) are presented in column 5 of Table~1
(lines 2 and 3, respectively).

In the end of this section, we consider the interaction of an atom and graphene
with $\Delta<2\mu$ in the limiting case of $\Delta=0$. This case was not
investigated in the literature so far because \nur(\ref{eq77}) and (\ref{eq80})
were derived \cite{148} under a condition $\Delta>\hbar\omega_c$.

{}From the first formula of (\ref{eq30}) we have
\begin{equation}
\Pi_{00,0}(\kb,0,0,\mu)\approx\frac{8\alpha c\mu}{v_F^2}\equiv Q_0\, .
\label{eq83}
\end{equation}
\noindent
Then, from \nur(\ref{eq75}) and (\ref{eq79}), one finds
\begin{equation}
\dT\rM(0,\kb)\approx
\frac{2\hbar \kb\dT\Pi_{00,0}(\kb,T,0,\mu)}{Q_0^2}\,,
\qquad
\dT\rE(0,\kb)\approx
-\frac{\dT\Pi_{0}(\kb,T,0,\mu)}{2\hbar k_{\bot}^3}\,.
\label{eq84}
\end{equation}
\noindent
Here, $\dT\Pi_{00,0}(\kb,T,0,\mu)$ and  $\dT\Pi_{0}(\kb,T,0,\mu)$ are
defined by \nur(\ref{eq76}) and (\ref{eq78a}), respectively, with
$l=0$ and $\Delta=0$.

Using these definitions, we obtain
\begin{equation}
\dT\Pi_{00,0}(\kb,T,0,\mu)\approx
\frac{\alpha\pi\kb\hbar c}{v_F}e^{-\frac{\mu}{k_BT}},
\qquad
\dT\Pi_{0}(\kb,T,0,\mu)\approx
\frac{\alpha\pi v_F k_{\bot}^3\hbar}{c}e^{-\frac{\mu}{k_BT}}.
\label{eq86}
\end{equation}
\noindent
Substituting these expressions to the term of \ur(\ref{eq41}) with $l=0$ and integrating
with respect to $\kb$, one arrives at
\begin{equation}
\dbT{\cal F}_{CP}(a,T)\sim -\frac{\alpha_0k_BT}{a^3}
e^{-\frac{\mu}{k_BT}}.
\label{eq87}
\end{equation}
\noindent
{}From the comparison with (\ref{eq77}), it is seen that, although this thermal
correction has the same functional form as (\ref{eq77}) obtained for
$\Delta>\hbar\omega_c$, it has the opposite sign.

Next, we substitute \nur(\ref{eq72}), (\ref{eq75}), (\ref{eq76}), (\ref{eq78a}), and
(\ref{eq79})  in the sum of all terms of (\ref{eq41}) with $l\geqslant 1$.
In this case the integration with respect of $\kb$ and a summation with respect to
$l$ result in
\begin{equation}
\dcT{\cal F}_{CP}(a,T)\sim -\frac{\alpha_0\hbar c}{a^4}
e^{-\frac{\mu}{k_BT}}.
\label{eq88}
\end{equation}
\noindent
This is the same functional dependence on $T$ as in \ur(\ref{eq80}) obtained for
$\Delta>\hbar\omega_c$, but again of the opposite sign. The results
(\ref{eq87}) and (\ref{eq88}) are quite expected because for $\Delta<2\mu$
the low-temperature dependence of the thermal correction $\deT{\cal F}_{CP}$
is mostly determined by the chemical potential.

The implicit thermal correction  $\daT{\cal F}_{CP}$ in the case $\mu\neq 0$,
$\Delta=0$ is given by (\ref{eq73}) because the condition  $\Delta>\hbar\omega_c$
was not used in the derivation of this equation. By comparing \nur(\ref{eq73}),
(\ref{eq87}), and (\ref{eq88}), we conclude that in the case $\mu\neq 0$,
$\Delta=0$
the low-temperature behavior of the total thermal correction to the
Casimir-Polder energy is given by
\begin{equation}
\dT{\cal F}_{CP}(a,T)\sim\daT{\cal F}_{CP}(a,T)\sim
-\frac{\alpha_0\mu(k_BT)^2}{(\hbar c)^2a}.
\label{eq89}
\end{equation}

The respective Casimir-Polder entropy
\begin{equation}
S_{CP}(a,T)\sim
\frac{\alpha_0\mu k_B^2T}{(\hbar c)^2a}
\label{eq90}
\end{equation}
\noindent
vanishes with vanishing temperature, i.e., the Nernst heat theorem
is satisfied.

%%%%%%%%%%%%%%%%%%%%%%%%%%%%%%%%%%%%%%%%%%%%%%%%%%%%%%%%%%%%%%%%%%%%%%%%%%%%
\section{Low-Temperature Behavior of the Casimir-Polder Free Energy and Entropy
for Two Graphene Sheets Possessing the Energy Gap and Chemical Potential}

In this section, we consider the same configuration as in Section~7, i.e.,
two parallel graphene sheets, but assume that they are characterized by some
values of $\Delta$ and $\mu$. The Casimir free energy per unit area of the
graphene sheets is given by \ur(\ref{eq1}), the reflection coefficients are
expressed by (\ref{eq32}), and the explicit expressions for the polarization
tensor entering (\ref{eq32}) are presented in (\ref{eq26})--(\ref{eq29}).
A behavior of the Casimir free energy at low $T$ is derived along the same
lines as in Sections~7 and 8.
Specifically, the asymptotic expansions are made using the small parameters
presented in \nur(\ref{eq68a}) and (\ref{eq82a}).
For this reason, below we concentrate only
on the main results and consider one new case.

As usual, we start from graphene satisfying the condition $\Delta>2\mu$.
Under this condition the quantities
$\Pi_{00,l}^{(0)}$ and $\Pi_{l}^{(0)}$ in (\ref{eq27}) have the meaning of
the polarization tensor components at $T=0$ whereas $\Pi_{00,l}^{(1)}$
and $\Pi_{l}^{(1)}$ in (\ref{eq29}) are the thermal corrections to them.
The thermal correction to the Casimir energy is defined  in (\ref{eq49})
and (\ref{eq50}). It is again presented as a sum of the implicit and
explicit contributions in (\ref{eq51})--(\ref{eq53}).

The low-temperature behavior of the implicit contribution is found under
the condition $\Delta>\hbar\omega_c$. It is determined by the TE mode
of the electromagnetic field and is given by (see \cite{149} for details)
\begin{equation}
\daT{\cal F}_C(a,T)\sim -\frac{(k_BT)^5}{(\hbar c)^2\Delta^2}.
\label{eq91}
\end{equation}
\noindent
This is a different behavior from a pristine graphene where the result in
({\ref{eq54}) is determined by the TM mode.

To calculate the explicit thermal correction one can again use the identity
(\ref{eq55}) and in the first perturbation order in $\dT r_{\lambda}$
represent $\deT{\cal F}_C(a,T)$ by \ur(\ref{eq56}).
Then this correction is presented in (\ref{eq57}) as a sum of the
term of (\ref{eq56}) with $l=0$ and of all other terms of (\ref{eq56}) with
$l\geqslant 1$. It should be taken into account, however, that we can
restrict ourselves by the first perturbation order only in the case when
$r_{\lambda}^{(0)}\neq 0$. This is just the case of a pristine graphene
considered in Section~7 and for a real graphene sheets with  $\Delta>2\mu$
under consideration now. As for graphene with $\Delta<2\mu$ where
$\orE(0,\kb)=0$, the special care is needed in that case (see below).

Using the term of (\ref{eq56}) with $l=0$, under the condition
$k_BT\ll\Delta-2\mu$ it was obtained (see \cite{149} for details)
\begin{equation}
\dbT{\cal F}_C(a,T)\sim -\frac{(k_BT)^2}{a^2\Delta}\deE.
\label{eq92}
\end{equation}
\noindent
This result is determined by the contribution of the TM mode.

The behavior of the sum of all terms of (\ref{eq56}) with $l\geqslant 1$
at low temperature was found  under the conditions $\Delta>\hbar\omega_c$,
$k_BT\ll\Delta-2\mu$. The result determined by the TM mode is the
following \cite{149}:
\begin{equation}
\dcT{\cal F}_C(a,T)\sim -\frac{k_BT}{a^2}\deE.
\label{eq93}
\end{equation}
%\noindent

Comparing \nur(\ref{eq91})--(\ref{eq93}), one concludes that for two parallel
graphene sheets satisfying the condition $\Delta>2\mu$ the behavior of the
total thermal correction to the Casimir energy is given by
\begin{equation}
\dT{\cal F}_C(a,T)\sim
\daT{\cal F}_C(a,T)\sim -\frac{(k_BT)^5}{(\hbar c)^2\Delta^2}.
\label{eq94}
\end{equation}
\noindent
The respective Casimir entropy
\begin{equation}
S_C(a,T)\sim \frac{k_B(k_BT)^4}{(\hbar c)^2\Delta^2}
\label{eq95}
\end{equation}
\noindent
goes to zero when $T$ goes to zero in accordance to the Nernst heat theorem.

The asymptotic behaviors (\ref{eq94}) and (\ref{eq95}) obtained for the case
$\Delta>2\mu$ are presented in column 3 of Table~2 (lines 2 and 3, respectively).

Similar to the Casimir-Polder interaction, considered in Section~8, the
configuration of two parallel graphene sheets with $\Delta=2\mu$ can be considered
as the limiting case of two sheets with $\Delta>2\mu$. As a result, \ur(\ref{eq91})
remains valid and \nur(\ref{eq92}) and (\ref{eq93}) are replaced with
\begin{equation}
\dbT{\cal F}_C(a,T)\sim -\frac{(k_BT)^2}{a^2\Delta},
\qquad
\dcT{\cal F}_C(a,T)\sim -\frac{k_BT}{a^2}.
\label{eq96}
\end{equation}
%\noindent

Comparing \nur(\ref{eq91}) and (\ref{eq96}), we conclude that for two
graphene sheets with $\Delta=2\mu$ the
total thermal correction to the Casimir energy behaves at low $T$ as
\begin{equation}
\dT{\cal F}_C(a,T)\sim
\deT{\cal F}_C(a,T)\sim -\frac{k_BT}{a^2}
\label{eq97}
\end{equation}
\noindent
and is determined by the TM mode in an explicit contribution.

The respective Casimir entropy
\begin{equation}
S_C(a,T)\sim \frac{k_B}{a^2}
\label{eq98}
\end{equation}
\noindent
does not vanish at $T=0$ and, thus, violates the Nernst heat theorem
(see the next section for a discussion of this result).

The asymptotic behaviors (\ref{eq97}) and (\ref{eq98}) valid for
$\Delta=2\mu$ are included in column 4 of Table~2 (lines 2 and 3, respectively).

We also note that, similar to Section~8, \nur(\ref{eq91})--(\ref{eq95}) are
applicable in the case $\mu=0$, whereas \nur(\ref{eq96})--(\ref{eq98}) are
valid only for $\Delta=2\mu\neq 0$.

The last remaining case is two graphene sheets with $\Delta<2\mu$. Here, the
implicit thermal correction is defined in (\ref{eq52}) where the reflection
coefficients are given in (\ref{eq72}). The low-temperature behavior of
$\daT{\cal F}_C$ is found under the conditions
$\sqrt{4\mu^2-\Delta^2}>\hbar\omega_c$ and $k_BT\ll\hbar\omega_c$ by using
the Abel-Plana formula (\ref{eq43}) with the following result determined by
the TE mode \cite{149}:
\begin{equation}
\daT{\cal F}_C(a,T)\sim -\frac{a(4\mu^2-\Delta^2)(k_BT)^2}{(\hbar c)^3}.
\label{eq99}
\end{equation}
\noindent
This result is evidently inapplicable at $\mu=0$, but $\Delta=0$ is allowed.

The contribution of the TM mode to $\dbT{\cal F}_{C}$ can be
calculated using (\ref{eq56}) because $\Pi_{00,0}(\kb,0,\Delta,\mu)\neq 0$
and, as a consequence, $\orM(0,\kb)\neq 0$. Then, under the conditions
$\sqrt{4\mu^2-\Delta^2}>\hbar\omega_c$, $\Delta>\hbar\omega_c$, and
$k_BT\ll 2\mu-\Delta$ the behavior of $\dbT{\cal F}_C(a,T)$ is found by
substituting \nur(\ref{eq75}) and (\ref{eq76}) in (\ref{eq56}) with
the result (see \cite{149} for details)
\begin{equation}
\dbT{\cal F}_{C,{\rm TM}}(a,T)\sim \frac{\hbar c\Delta k_BT}{a^3\mu^2}\meE.
\label{eq100}
\end{equation}

In order to find the low-temperature behavior of $\dbT{\cal F}_{C,{\rm TE}}$,
one should refuse from using the perturbation expansion in the parameter
$\dT\Pi_0/\Pi_0(\kb,0,\Delta,\mu)$ and consider the term of (\ref{eq53})
with $l=0$ taking into account that $\orE(0,\kb)=0$ and, thus,
$\rE(0,\kb)=\dT\rE(0,\kb)$:
\begin{eqnarray}
\dbT{\cal F}_{C,{\rm TE}}(a,T)&=&\frac{k_BT}{4\pi}
\int_{0}^{\infty}\!\!\!\kb d\kb\ln\left\{1-
\left[\dT\rE(0,\kb)\right]^2e^{-2a\kb}\right\}
\nonumber \\
&\approx& -\frac{k_BT}{4\pi}
\int_{0}^{\infty}\!\!\!\kb d\kb
\left[\dT\rE(0,\kb)\right]^2e^{-2a\kb}.
\label{eq101}
\end{eqnarray}
\noindent
The low-temperature behavior of (\ref{eq101}) was found in \cite{149}.
It was shown that it is much smaller in magnitude than
$\dbT{\cal F}_{C,{\rm TM}}$ and, thus, the behavior of
$\dbT{\cal F}_{C}$ is given by \ur({\ref{eq100}).

The last quantity contributing to the thermal correction is $\dcT{\cal F}_{C}$.
For $l\geqslant 1$ we have $\Pi_{00,l}(\kb,0,\Delta,\mu)\neq 0$ and
$\Pi_l(\kb,0,\Delta,\mu)\neq 0$, so that this quantity at low $T$ can be estimated
using the sum of terms in \ur(\ref{eq56}) with $l\geqslant 1$.
It was shown \cite{149} that, depending on the values of parameters, the TM and TE
modes can give the same order contributions to $\dcT{\cal F}_{C}$ which are
the following:
\begin{equation}
\dcT{\cal F}_{C}(a,T)\sim \frac{(\hbar c)^2}{a^4}\left(
\frac{\Delta}{\alpha\mu^2}+\frac{\alpha^2\hbar c\sqrt{4\mu^2-\Delta^2}}{a\Delta^3}
\right)\,\meE.
\label{eq102}
\end{equation}
\noindent
Here, we have also included the dimensionless fine structure constant $\alpha$
because, in contrast to all the above expressions, it cannot be separated as a common
factor.

{}From the comparison of \nur(\ref{eq99}), (\ref{eq100}), and (\ref{eq102}),
it is seen that in the case $\Delta<2\mu$ the total thermal correction to
the Casimir energy behaves at low $T$ as
\begin{equation}
\dT{\cal F}_C(a,T)\sim
\daT{\cal F}_C(a,T)\sim -\frac{a(4\mu^2-\Delta^2)(k_BT)^2}{(\hbar c)^3}.
\label{eq103}
\end{equation}
\noindent
This leads to the following low-temperature behavior of the Casimir entropy:
\begin{equation}
S_C(a,T)\sim \frac{a(4\mu^2-\Delta^2)k_B^2T}{(\hbar c)^3}.
\label{eq104}
\end{equation}
\noindent
With vanishing $T$, the Casimir entropy vanishes in accordance with the
Nernst heat theorem.

The asymptotic behaviors (\ref{eq103}) and (\ref{eq104}) applicable at
$\Delta<2\mu$ (including $\Delta=0$, see below) are presented in column 5
of Table~2 (lines 2 and 3, respectively).

To conclude this section, we consider the low-temperature behavior of the
thermal correction to the Casimir energy for the case $\mu\neq 0$, $\Delta=0$
which was not considered in the literature up to the present.
As to \ur(\ref{eq99}) for the implicit contribution to the thermal correction,
it remains valid in the case $\Delta=0$ because the condition $\Delta>\hbar\omega_c$
was not used in its derivation. Because of this, below we concentrate on the
explicit contribution to the thermal correction.

We start with the part $\dbT{\cal F}$ of this contribution. When considering the
TM mode, one can use the term with $l=0$ in the perturbative \ur(\ref{eq56}).
Here, the reflection coefficient $\orM(0,\kb)\approx 1$. This result follows
from the first formula in (\ref{eq72}) taken at $\xi_0=0$ and (\ref{eq83}) under
the condition $2\mu>\hbar\omega_c$. The thermal correction to the reflection
coefficient $\dT\rM(0,\kb)$, also entering (\ref{eq56}), is given by the first
formulas in (\ref{eq84}) and  (\ref{eq86}). Substituting these in the term of
(\ref{eq56}) with $l=0$, one obtains
\begin{equation}
\dbT{\cal F}_{C,{\rm TM}}(a,T)\sim -\frac{(\hbar c)^2k_BT}{a^4\mu^2}
e^{-\frac{\mu}{k_BT}}.
\label{eq105}
\end{equation}

In order to find the behavior of $\dbT{\cal F}_{C,{\rm TE}}$ at low $T$,
one should use \ur(\ref{eq101}) where $\dT\rE$ is defined in the second
formulas of (\ref{eq84}) and  (\ref{eq86}). In this way, the factor of
$\exp[-2\mu/(k_BT)]$ is obtained, i.e., an additional exponentially small
multiple as compared to (\ref{eq105}). This means that \ur(\ref{eq105})
in fact describes a behavior of the full contribution
$\dbT{\cal F}_{C}(a,T)$ to the thermal correction at low $T$.

It remains to consider the contribution $\dcT{\cal F}_C$ to the thermal correction.
This can be done in the same way as in Section~8. As in all cases considered
above, a summation over all nonzero Matsubara frequencies adds the factor
$1/\tau$ to the low-temperature behavior of $\dbT{\cal F}_C$.
As a result one obtains
\begin{equation}
\dcT{\cal F}_{C}(a,T)\sim -\frac{(\hbar c)^3}{a^5\mu^2}
e^{-\frac{\mu}{k_BT}}.
\label{eq106}
\end{equation}

{}From  \nur(\ref{eq99}), (\ref{eq105}), and (\ref{eq106}) we conclude that
 the total thermal correction to the Casimir energy of two graphene sheets
with $\mu\neq 0$ and $\Delta=0$ is given by
\begin{equation}
\dT{\cal F}_C(a,T)\sim
\daT{\cal F}_C(a,T)\sim -\frac{a\mu^2(k_BT)^2}{(\hbar c)^3}.
\label{eq107}
\end{equation}
\noindent
The respective Casimir entropy
\begin{equation}
S_C(a,T)\sim \frac{a\mu^2k_B^2T}{(\hbar c)^3}
\label{eq108}
\end{equation}
\noindent
goes to zero with vanishing  $T$ as is demanded by the Nernst heat theorem.

%%%%%%%%%%%%%%%%%%%%%%%%%%%%%%%%%%%%%%%%%%%%%%%%%%%%%%%%%%%%%%%%%%%%%%%%
\section{Discussion: How Graphene May Point the Way to Resolution of
Thermodynamic Problems in the Lifshitz Theory}

As is seen from the foregoing, the Lifshitz theory of the Casimir-Polder
interaction of an atom with a pristine graphene sheet and the Casimir
interaction between two sheets of pristine graphene is consistent with
the requirements of thermodynamics. This is apparent from the fact that
the Casimir-Polder and Casimir entropies go to zero with  vanishing
temperature in accordance with the Nernst heat theorem if the dielectric
response of a pristine graphene is described on the basis of first
principles of thermal quantum field theory by means of the polarization
tensor.

A comparison between the case of a graphene sheet and a metallic plate
raises several questions. As mentioned in Section~1 and discussed in
Section~3, the Lifshitz theory of the Casimir interaction between two
metallic plates with perfect crystal lattices violates the Nernst heat
theorem if the dielectric response of a metal takes into account the
relaxation properties of free electrons. An agreement with
thermodynamics is restored if the dissipation properties of conduction
electrons are omitted in calculations. Graphene can be considered as
a conductor because its electrical conductivity remains nonzero
at zero temperature \cite{49,50,51}, and the polarization tensor results
in a complex effective dielectric permittivity of graphene taking the
relaxation properties into account. Nevertheless, for a pristine
graphene sheets, the Lifshitz theory is consistent with the Nernst heat
theorem and with the measurement data, which is not the case when the
same theory is applied to the Casimir interaction between two metallic
plates taking the proper account of the relaxation properties of free
electrons.

For real graphene sheets possessing some nonzero energy gap $\Delta$
and chemical potential $\mu$ the situation is largely the same.
According to the results presented above, the Lifshitz theory of both
the Casimir-Polder and Casimir interactions involving real graphene
sheets with $\Delta>2\mu$ or $\Delta<2\mu$ satisfies the Nernst heat
theorem. For real graphene the Nernst heat
theorem is violated in the only case of an exact equality
$\Delta=2\mu$ (see Sections 8 and 9). In this case both the Casimir-Polder
and Casimir entropies at zero temperature take nonzero values depending on
the parameters of a system. One should note, however, that the values
of $\Delta$ and $\mu$ for a specific graphene sample cannot be known
exactly. Thus, from the practical standpoint, the case of graphene sheets
with an exact equality $\Delta=2\mu$ should be considered as a singular one
and physically unrealizable. It is interesting to note also that the
real part of electrical conductivity of graphene as a function of frequency
undergoes a qualitative change when the conditions $\Delta>2\mu$ and
$\Delta<2\mu$ substitute for one another \cite{114}.

It should be also noted that the polarization tensor considered
above in Section~5 is obtained without external magnetic field and leads
to magnetic permeability of graphene equal to unity. An actual graphene
sheet, however, exhibits diamagnetism \cite{49,50}, i.e., its magnetic
permeability is slightly less than unity. Although for both metallic and
dielectric materials it was already demonstrated that an account of
magnetic properties has no effect on the satisfaction or violation of
the Nernst heat theorem \cite{37,46,48},  which are determined entirely
by the dielectric permittivity, for graphene this question was not considered
so far and deserves further attention.

Thus,  with this reservation, one can argue that
for both ideal (pristine) and real (gapped and doped) graphene
sheets described by the polarization tensor the Lifshitz theory of the
Casimir-Polder and Casimir interactions is thermodynamically consistent
in all physically realizable cases.
The question arises why there are so significant differences in application
of the Lifshitz theory to graphene, on the one hand, and to ordinary metals
and dielectrics, on the other hand.

It was hypothesized \cite{149} that this problem has roots stretching back
to the electrodynamics of solids. As mentioned in Section~3, a response of
metals to the low-frequency electromagnetic fields is usually described by
the dissipative Drude model. Although it is impossible to derive this model
starting from the first principles of thermal quantum field theory,
it does have a great number of confirmations in the areas of both electrical
and optical phenomena involving electromagnetic fields on the mass shell.
In order to calculate the Casimir force, however, it is necessary to know
the response function to both the propagating waves (on the mass shell) and
to the evanescent ones (off the mass shell). In the latter case, there are
no direct experimental evidences in favor of the Drude model and some
doubts are cast upon its validity. As to graphene, its response function
is derived on the basis of first principles of thermal quantum field
theory. It is equally applicable to describe the dielectric response of
graphene to both propagating and evanescent fields. This may explain why
the Lifshitz theory leads to contradictions with the Nernst heat theorem
for metals described by the Drude model, but is thermodynamically
consistent for graphene systems.

In such a manner graphene may point the way for resolution of the problem
in Casimir physics which is often called the Casimir puzzle or the
Casimir conundrum.

%%%%%%%%%%%%%%%%%%%%%%%%%%%%%%%%%%%%%%%%%%%%%%%%%%%%%%%%%%%%%%%%%%%%%%%%%
\section{Conclusions}

To summarize, we can say that the Lifshitz theory provides a reliable foundation
for a theoretical description of the Casimir-Polder and Casimir interactions.
Being originally formulated for the case of material semispaces described by the
frequency-dependent dielectric permittivities, it is now generalized for any
configurations with the known reflectivity properties including graphene.

An important conclusion is that the Lifshitz theory of the Casimir-Polder and
Casimir interactions in graphene systems is consistent with the requirements of
thermodynamics in all physically realizable cases if the dielectric response
of graphene is described on the basis of first principles of thermal quantum field
theory using the polarization tensor in (2+1)-dimensional space-time.

It should be pointed out that the derivation of the polarization tensor at any
temperature not only provided a full theoretical description of the electromagnetic
response of graphene, but has also stimulated development of other approaches
using the density-density correlation functions and the electrical conductivities of
graphene as basic quantities.

One more conclusion is that the thermodynamic properties of the Casimir and
Casimir-Polder interactions in graphene systems reviewed and derived above provide
the basis for a resolution of outstanding problems in the Casimir physics, which
remain unresolved for the last twenty years and are detrimental for its further
development and applications.

%%%%%%%%%%%%%%%%%%%%%%%%%%%%%%%%%%%%%%%%%%
%\authorcontributions{Investigation, both authors;  writing, both authors.}

%%%%%%%%%%%%%%%%%%%%%%%%%%%%%%%%%%%%%%%%%%
\funding{V.M.M.~was partially funded by the Russian Foundation for Basic Research grant number
19-02-00453 A.}

%%%%%%%%%%%%%%%%%%%%%%%%%%%%%%%%%%%%%%%%%%
\acknowledgments{This work was partially supported by the Peter
the Great Saint Petersburg Polytechnic University in the framework
of the Program ``5--100--2020".
V.M.M.\ was partially supported by the Russian Government Program of Competitive Growth of Kazan Federal University.}

%%%%%%%%%%%%%%%%%%%%%%%%%%%%%%%%%%%%%%%%%%
%\conflictsofinterest{The authors declare no conflict of interest.}

%%%%%%%%%%%%%%%%%%%%%%%%%%%%%%%%%%%%%%%%%%
\reftitle{References}

%%%%%%%%%%%%%%%%%%%%%%%%%%%%%%%%%%%%%
\end{document}